\documentclass[journal]{IEEEtran}
\usepackage[latin9]{inputenc}
\usepackage{geometry}
\usepackage{color}
\usepackage{array}
\usepackage{float}
\usepackage{amsmath}
\usepackage{amssymb}
\usepackage{graphicx}

\makeatletter

\providecommand{\tabularnewline}{\\}
\floatstyle{ruled}
\newfloat{algorithm}{tbp}{loa}
\providecommand{\algorithmname}{Algorithm}
\floatname{algorithm}{\protect\algorithmname}

\usepackage{amsfonts}
\usepackage{mathrsfs}
\usepackage{mathrsfs}\usepackage[noblocks]{authblk}
\usepackage[compress]{cite}
\usepackage{bm}
\usepackage{algorithm}
\usepackage{algorithmic}
\usepackage{epsfig}\usepackage{graphics}\usepackage{subfigure}\usepackage{epsfig}\usepackage{epstopdf}
\usepackage{graphics}\usepackage{subfigure}\usepackage{upgreek}\usepackage{caption}\usepackage{theorem}
\usepackage{breqn}
\usepackage{multicol}
\usepackage{eufrak}
\usepackage{eucal}
\usepackage{array}
\usepackage[compress]{cite}
\usepackage{algorithm}
\usepackage{algorithmic}
\allowdisplaybreaks[4]

\newtheorem{lemma}{Lemma}\newtheorem{proposition}{Proposition}\theoremheaderfont{\normalfont\bfseries}

\makeatother

\begin{document}
\title{Rate-Splitting Multiple Access for Multi-antenna Downlink Communication
Systems: Spectral and Energy Efficiency Tradeoff}
\author{Gui~Zhou, Yijie Mao,~\IEEEmembership{Member,~IEEE}, Bruno~Clerckx,~\IEEEmembership{Senior~Member,~IEEE}
\thanks{G. Zhou is with the School of Electronic Engineering and Computer
Science at Queen Mary University of London, London E1 4NS, U.K. (e-mail:
g.zhou@qmul.ac.uk). B. Clerckx and Y. Mao are with the Electrical
and Electronic Engineering Department, Imperial College London, London
SW7 2AZ, U.K. (e-mail: {b.clerckx, y.mao16}@imperial.ac.uk). } }
\maketitle
\begin{abstract}
Rate-splitting (RS) has recently been recognized as a promising physical-layer
technique for multi-antenna broadcast channels (BC). Due to its ability
to partially decode the interference and partially treat the remaining
interference as noise, RS is an enabler for a powerful multiple access,
namely rate-splitting multiple access (RSMA), that has been shown
to achieve higher spectral efficiency (SE) and energy efficiency (EE)
than both space division multiple access (SDMA) and non-orthogonal
multiple access (NOMA) in a wide range of user deployments and network
loads. As SE maximization and EE maximization are two conflicting
objectives in the moderate and high signal-to-noise ratio (SNR) regimes,
the study of the tradeoff between the two criteria is of particular
interest. In this work, we address the SE-EE tradeoff by studying
the joint SE and EE maximization problem of RSMA in multiple input
single output (MISO) BC with rate-dependent circuit power consumption
at the transmitter. To tackle the challenges coming from multiple
objective functions and rate-dependent circuit power consumption,
we first propose two models to transform the original problem into
two single-objective problems, namely, weighted-sum method and weighted-power
method. A low-complexity algorithm with closed-form
solution is proposed to solve each single-objective problem in the
two-user system. For the generalized $K$-user system, a successive
convex approximation (SCA)-based algorithm is then proposed to optimize
the precoders of each transformed problem. Numerical results show
that our algorithm converges much faster than existing algorithms.
In addition, the performance of RSMA is superior to or equal to SDMA
and NOMA in terms of both SE and EE and their tradeoff. 
\end{abstract}


\section{Introduction}

Rate-splitting (RS) has recently emerged in multi-antenna broadcast
channel (BC) as a powerful and robust non-orthogonal transmission
technique and interference management strategy for cellular networks.
At the base station (BS), the message intended for each user is split
into a common and a private part. After jointly encoding the common
parts into a common stream to be decoded by all users and independently
encoding the private parts into the private streams for the corresponding
users only, BS linearly precodes all the encoded streams and broadcasts
the superimposed data streams to all users. By allowing each user
to sequentially decode the common stream and the intended private
stream with the assistance of successive interference cancellation
(SIC), RS grants all users the ability to partially decode the interference
and partially treat the remaining interference as noise.

The study of RS dates back to early 1980s in \cite{RS-TIT1981} for
the two-user single-input single-output (SISO) interference channel
(IC). The flexibility of RS in dealing with interference in SISO IC
has motivated the investigation of the benefits of RS in modern multiple
input single output (MISO) BC \cite{Bruno-2016-RS,Bruno2019}. As
the two fundamental indicators for a communication system design,
spectral efficiency (SE) demonstrates the amount of information to
be transmitted per unit of time while energy efficiency (EE) demonstrates
how much information rate can be transmitted per unit of energy. Existing
literature of RS in MISO BC has shown that RS is an enabler for a
powerful multiple access, namely rate-splitting multiple access (RSMA),
that achieves both higher SE and EE over space division multiple access
(SDMA) and non-orthogonal multiple access (NOMA) in both perfect Channel
State Information at the Transmitter (CSIT) \cite{Mao2019EE,Mao2018journal,CRAN2019ACCESS,Mao2019icc}
and imperfect CSIT \cite{Hao2015RS,Dai2016massive,Dai2017hybrid,Joudeh2016Robust,Joudeh2016Tcom,Joudeh2017Rate,mao2020tcom}.
Compared with RSMA that dynamically partially decodes the interference
and partially treats interference as noise, SDMA and NOMA fall into
two extreme interference management cases where users in SDMA always
decode their intended signal by fully treating any residual interference
as noise and users in NOMA always fully decode the interference generated
by the users with weaker channel strengths \cite{Mao2018journal}.
Besides conventional MISO BC, the benefits of RSMA in the SE domain
have been further demonstrated in massive MISO system \cite{Dai2016massive},
millimeter wave system \cite{Dai2017hybrid}, overloaded system \cite{Joudeh2017Rate},
MISO BC with user relaying \cite{zhangjian-letter,mao2020twc}, etc.
The benefits of RSMA from an EE perspective are also investigated
in MISO BC \cite{Mao2018Energy}, and MISO BC with a common message
(so-called non-orthogonal unicast and multicast transmission) \cite{Mao2019EE}.

The results in the existing literature \cite{liye2011,Shiwen2013letter}
clarify that both EE and SE increase with the signal-to-noise ratio
(SNR) in the low SNR regime. Therefore, both EE and SE are maximized
by consuming all transmit power. However, in the moderate and high
SNR regimes, SE still increases with SNR while EE starts decreasing
with SNR. In another words, SE is maximized by making use of the full
available transmit power, while this is not the case for EE where
transmitting using a portion of the available transmit power budget
is preferred to maximize EE. Therefore, EE and SE conflict with each
other in the moderate and high SNR regimes leading to a fundamental
tradeoff between SE and EE.

In recent years, the existence of the SE-EE tradeoff is well recognized
in the literature, as shown in \cite{tradeoff-2,tradeoff2016tvt,tradeoff2018twc}.
In some cases, the tradeoff is achieved by optimizing the EE metric
constrained by the SE and transmit power constraints \cite{tradeoff-2}.
The higher the SE target, the lower the EE generated. However, such
method requires a given SE constraint and the optimization problem
comes with a feasibility issue. Whether the given SE constraint can
be met requires extra feasibility testing. Jointly optimizing SE and
EE gets rid of those issues and it is a more flexible and general
method to deal with the SE-EE tradeoff problem \cite{tradeoff2016tvt,tradeoff2018twc}.

The SE-EE tradeoff optimization problem is a multi-objective optimization
(MOO) problem which is generally transformed to its corresponding
single-objective optimization (SOO) problem by adopting the weighted
sum method \cite{tradeoff2016tvt,Tervo2018tradeoff}. Two widely used
precoder design frameworks to solve the transformed SOO problem are
Dinkelbach's framework \cite{FP-Dinkelbach}\cite{Shiwen2013letter}
and successive convex approximation (SCA) framework \cite{Razaviyayn2014PHD}\cite{Oskari2018}
or called inner approximation (IA) framework \cite{Nguyen2019EE}.
The key step of the Dinkelbach's framework is to transform the fractional
program into a sequence of parametric problems by introducing an auxiliary
variable, which can then be solved by zero forcing (ZF) \cite{ZF-1},
weighted minimum mean square error (WMMSE) \cite{WMMSE-1}, SCA methods
\cite{7437396}, monotonic optimization \cite{Zappone2017EE,Matthiesen2019EE}.
It is indeed a two-layer iterative algorithm framework, which optimizes
the parameter in the outer layer and precoder in the inner layer.
In comparison, SCA framework can be directly applied to solve the
SE-EE tradeoff problem by approximating the fractional EE metric as
well as other non-convex expressions into their convex approximation
counterparts, which results in a one-layer iterative algorithm. The
SCA framework has shown its performance advantages in terms of convergence
compared to the Dinkelbach's framework in conventional communication
systems with orthogonal multiple access (OMA) \cite{Oskari2018}.

In this work, we investigate the SE-EE tradeoff of RSMA in multi-user
multi-antenna systems. The major contributions of this work are as
follows: 
\begin{itemize}
\item We investigate the SE-EE tradeoff achieved by RSMA in a MISO BC. Previous
works on RSMA either address the SE (as in \cite{Mao2019EE,Mao2018journal,CRAN2019ACCESS,Mao2019icc,Hao2015RS,Dai2016massive,Dai2017hybrid,Joudeh2016Robust,Joudeh2016Tcom,Joudeh2017Rate})
or EE (as in \cite{Mao2018Energy,Mao2019EE}), but the tradeoff between
SE and EE has never been studied. It should be reminded that, since
RSMA is a general framework for non-orthogonal transmission that subsumes
SDMA, NOMA, OMA and multicasting as special cases \cite{Bruno2019,Mao2018journal},
identifying the SE-EE tradeoff of RSMA automatically solves the SE-EE
tradeoff of those particular strategies. 
\item We formulate a MOO problem that jointly maximizes SE and EE of RSMA
in a MISO BC subject to a transmit power constraint. To obtain reasonable
operating points on the Pareto boundary that balance SE and EE, we
adopt two different approaches to convert the MOO problem into a SOO
problem, namely the weighted-sum and the weighted-power approaches.
The former transformation is achieved by maximizing the weighted sum
of SE and EE and the latter is to minimize the weighted sum of the
inverse of both SE and EE. 
\item To better characterize the SNR regime where SE and
EE have conflicting tendencies as SNR increases, we first propose
a low-complexity linear precoding algorithm for the two-user case
and analytically study the optimal transmit power to maximize the
SE and EE tradeoff. Motivated by the insights obtained from the two-user
case, we investigate the optimal precoding method to maximize the
SE-EE tradeoff in the generalized $K$-user case. Due to the non-convexity
of the transformed SOO problems, we propose a SCA-based framework
to solve both weighted-sum and weighted-power problems with two different
rate lower bounds to relax the non-convex rate constraints, namely
``LB I'' and ``LB II''. LB I is achieved by exploiting the convexity
of log function and directly using the first-order Taylor approximation
of the rate function. In comparison, LB II is obtained by approximating
the fractional signal-to-interference-plus-noise ratio (SINR) expression
only. The proposed SCA-based algorithm is shown to not only solve
the SE-EE tradeoff problem but also the individual SE and EE problems. 
\item We demonstrate through numerical results that RSMA achieves a larger
achievable SE and EE tradeoff regime than the conventional multi-user
SDMA and NOMA in different user deployments, SE and EE weights, transmit
power constraints, etc. The proposed low-complexity
precoding algorithm, though with a much lower computational complexity,
achieves almost the same trade-off performance as the proposed SCA
sub-optimal precoding algorithm. We further evaluate the convergence
and complexity of the developed algorithm by observing the CPU time
and the number of required iterations. The proposed SCA-based algorithm
with both lower bound LB I and II converges within a few iterations
and is faster than the existing Dinkelbach's algorithm. LB I converges
slightly slower than LB II but both converge to similar boundary point.
We conclude that RSMA can not only improve individual SE and EE, but
also achieve a better SE-EE tradeoff in multi-antenna BC. 
\end{itemize}
The remaining of this paper is organized as follows. In Section II,
the system model is specified and the optimization problems are formulated.
The weighted-sum approach is proposed to design the precoder in Section
III. A weighted-power approach is proposed to design the precoder
in Section IV. Section V and Section VI show the numerical results
and conclusion, respectively.

\noindent {\textbf{Notations:}} Vectors and matrices are denoted
by boldface lowercase and uppercase letters, respectively. The symbols
${\mathbf{x}}^{*}$, ${\mathbf{x}}^{\mathrm{T}}$ and ${\mathbf{x}}^{\mathrm{H}}$
denote the conjugate, transpose, and Hermitian (conjugate transpose)
of vector ${\mathbf{x}}$, respectively. Additionally, the symbol
${\mathbb{C}}$ denotes complex field and ${\mathbb{R}}$ represents
real field. The symbol $||\mathbf{x}||_{2}$ denotes 2-norm of vector
$\mathbf{x}$ and the symbol $||\mathbf{X}||_{F}$ denotes Frobenius
norm of matrix $\mathbf{X}$. The symbols $\mathrm{Tr}\{\cdot\}$,
$\mathrm{Re}\{\cdot\}$, and $|\cdot|$ denote the trace, real part
and modulus, respectively.

\section{System Model and problem formulation}

In this section, we specify the system model of RSMA in MISO BC followed
by the formulated SE and EE tradeoff problem. The rate-dependent power
consumption model of the transmitter is also specified in this section.

\begin{figure*}
\centering \includegraphics[width=6.8in,height=2in]{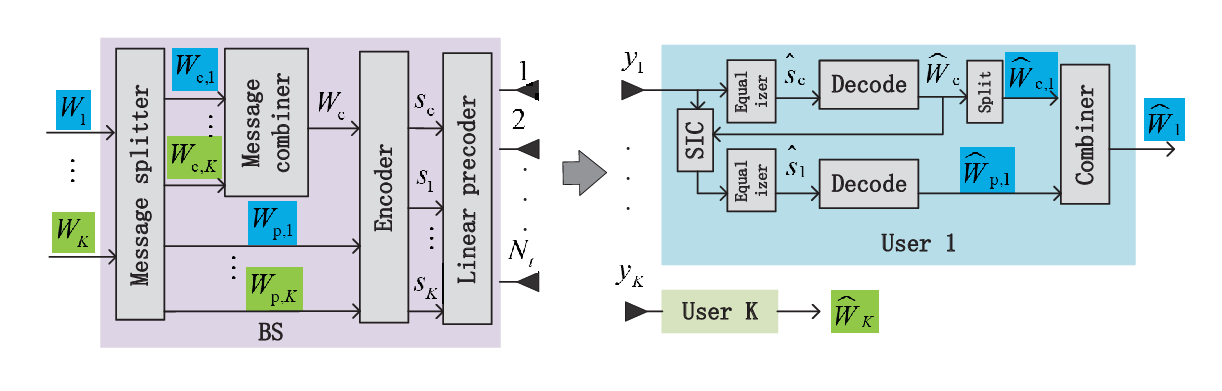} \caption{System architecture of RSMA for MISO BC.}
\label{rsma} 
\end{figure*}

\subsection{Rate-Splitting Transmit Signal Model}

In this work, we consider a downlink multi-antenna multi-user communication
system where one BS with $N_{t}$ transmit antennas transmits $K$
messages simultaneously to $K$ single-antenna users. The principle
of RSMA is shown in Fig. \ref{rsma}. In particular, at the BS, the
message $W_{k}$ intended for user $k$ is split into a private part
$W_{\mathrm{p},k}$ and a common part $W_{\mathrm{c},k}$. The private
parts $W_{\mathrm{p},1},\cdots,W_{\mathrm{p},K}$ are independently
encoded into the private streams $s_{1},\cdots,s_{K}$ and the common
parts of all users $W_{\mathrm{c},1},\cdots,W_{\mathrm{c},K}$ are
combined into a common message $W_{\mathrm{c}}$, which is encoded
into a common stream $s_{\mathrm{c}}$ using a public codebook. Denoting
$\mathbf{s}=[s_{\mathrm{c}},s_{1},\cdots,s_{K}]^{\mathrm{T}}$ and
assuming Gaussian signaling with $\mathbb{E}[\mathbf{s}\mathbf{s}^{\mathrm{H}}]=\mathbf{I}$,
the $K+1$ streams are linearly precoded by precoding vectors $\mathbf{f}_{\mathrm{c}},\mathbf{f}_{1},\cdots,\mathbf{f}_{K}$.
Define set $\mathcal{K}=\{1,2,...,K\}$ and set $\mathcal{K}_{\mathrm{c}}=\mathcal{K}\cup\{\mathrm{c}\}$,
the resulting transmit signal of RSMA can be written as 
\begin{align}
\mathbf{x}=\sum\limits _{i\in\mathcal{K}_{\mathrm{c}}}\mathbf{f}_{i}s_{i}.
\end{align}
We use a compact notation $\mathbf{F}$ to denote the family of the
precoder vectors as $\mathbf{F}=\left[\mathbf{f}_{\mathrm{c}},\mathbf{f}_{1},\cdots,\mathbf{f}_{K}\right]\in\mathbb{C}^{N_{\mathrm{t}}\times(K+1)}$.
It belongs to the power constraint set $\mathcal{S}=\{\mathbf{F}\left|||\mathbf{F}||_{F}^{2}\le\mathrm{P_{\mathrm{max}}}\right.\}$
where $\mathrm{P_{\mathrm{max}}}$ is the transmit power limit at
the BS.

The received signal of user $k$ is 
\begin{align}
y_{k}=\mathbf{h}_{k}^{\mathrm{H}}\sum\limits _{i\in\mathcal{K}_{\mathrm{c}}}\mathbf{f}_{i}s_{i}+n_{k},
\end{align}
where $\mathbf{h}_{k}\in\mathbb{C}^{N_{\mathrm{t}}\times1}$ is the
channel between BS and user $k$ and noise $n_{k}$ follows Gaussian
distribution, i.e., $\mathcal{N}(0,\sigma_{k}^{2})$. In this work,
we assume perfect CSIT\footnote{The investigation of the SE-EE tradeoff in RSMA under imperfect CSIT
will be considered in our future work.}.

At the receivers, each user firstly decodes the common stream by treating
all the private streams as noise, then SIC is used to remove the decoded
common stream from the received signal under the assumption of error-free
decoding\footnote{In this paper we use information theoretical rate expressions. It
means that the codeword lengths are long enough to apply Shannon theoretical
arguments and there is no decoding error. Since the obtained rates
are achievable, there is no error in decoding, and in the successive
interference cancellation process. In other words, perfect SIC is
not assumed for simplicity, but is a result of good channel codes.
The impact of finite block length and imperfect SIC on the system
performance is beyond the scope of this work, which will be our future
work.}. Each user then decodes its own private stream by treating other
private streams as noise. Thus, the spectral efficiencies (bit/s/Hz)
of the intended common and private streams at user $k$ are formulated
as 
\begin{align}
R_{\mathrm{c},k} & \left(\mathbf{F}\right)=\log_{2}\left(1+\frac{\mathbf{h}_{k}^{\mathrm{H}}\mathbf{f}_{\mathrm{c}}\mathbf{f}_{\mathrm{c}}^{\mathrm{H}}\mathbf{h}_{k}}{\sigma_{k}^{2}+\sum_{i\in\mathcal{K}}\mathbf{h}_{k}^{\mathrm{H}}\mathbf{f}_{i}\mathbf{f}_{i}^{\mathrm{H}}\mathbf{h}_{k}}\right),\label{common-rate-original}
\end{align}
\begin{align}
R_{k} & \left(\mathbf{F}\right)=\log_{2}\left(1+\frac{\mathbf{h}_{k}^{\mathrm{H}}\mathbf{f}_{k}\mathbf{f}_{k}^{\mathrm{H}}\mathbf{h}_{k}}{\sigma_{k}^{2}+\sum_{i\in\mathcal{K}\backslash\{k\}}\mathbf{h}_{k}^{\mathrm{H}}\mathbf{f}_{i}\mathbf{f}_{i}^{\mathrm{H}}\mathbf{h}_{k}}\right).\label{private-rate-original}
\end{align}
The achievable rate of the common stream is given as $R_{{\rm c}}\left(\mathbf{F}\right)=\min_{k\in\mathcal{K}}R_{\mathrm{c},k}\left(\mathbf{F}\right)$
to guarantee that all users are capable of decoding the common stream
successfully.

Finally, the system SE is the sum of the SE of the common and private
streams, which is given by 
\begin{equation}
f\left(\mathbf{F}\right)=\sum_{k\in\mathcal{\mathcal{K}_{\mathrm{c}}}}R_{k}\left(\mathbf{F}\right).\label{eq:sum-SE}
\end{equation}

\textbf{Remark 1}: Please keep in mind that the contribution of the
common stream is observed by comparing RSMA and traditional multiple
access techniques of SDMA and NOMA. Since this work considers sum
SE, we can see from (\ref{eq:sum-SE}) that splitting only one user's
message or multiple users' messages does not change the problem formulation
as well as its performance. Therefore, the split of a single user's
message is sufficient in this work.

\subsection{Spectral Efficiency Maximization}

The existing literature \cite{Dai2016massive,Dai2017hybrid,Joudeh2016Tcom,Joudeh2017Rate,Mao2018journal,Joudeh2016Robust,zhangjian-letter}
have investigated precoder design of RSMA for maximizing the SE in
various scenarios. The SE maximization problem is formulated as follows
\begin{align}
\mathop{\max}\limits _{\mathbf{F}\in\mathcal{S}} & \;\;f\left(\mathbf{F}\right).\label{RS-SE-problem}
\end{align}

Popular precoding techniques rely on closed form precoders based on
ZF method \cite{ZF-1} for private messages as in \cite{Bruno2019,Hao2015RS,Dai2016massive,Dai2017hybrid},
or optimized precoders based on convex optimization commonly relying
on an extended version of the WMMSE method \cite{WMMSE-1}, as in
\cite{Mao2018journal,Joudeh2016Robust,Joudeh2016Tcom,Joudeh2017Rate,zhangjian-letter}).

\subsection{Energy Efficiency Maximization}

As another important performance metric, EE benefits of RSMA in multi-antenna
BC have only been studied in a few literature \cite{Mao2018Energy,Mao2019EE}
and only constant circuit power consumption is considered in those
works. However, in practical communication systems, the circuit power
consumption contains two parts, namely the rate-independent fixed
part and the rate-dependent dynamic part \cite{Emil2015EE,Xiong2016ratedependent}.
The former is for basic circuit operations, e.g., channel estimation,
precoder chains, and linear processing at the BS while the latter
is for information processing, e.g., coding, decoding and backhaul
power consumption. In this work, we consider a more practical power
consumption model that has not been studied in the literature of RSMA
yet. We adopt the power consumption model of \cite{Xiong2016ratedependent,Isheden2010ratedependent},
where the rate-dependent circuit power consumption is written as a
linear function of the system sum-rate. Therefore, the total power
consumption is modeled as 
\begin{align}
g\left(\mathbf{F}\right)=||\mathbf{F}||_{F}^{2}+\mathrm{P_{c}}+\chi f\left(\mathbf{F}\right),\label{power}
\end{align}
where $||\mathbf{F}||_{F}^{2}$ is the transmit power consumption,
$\mathrm{P_{c}}$ is a constant representing the rate-independent
fixed power consumption, and $\chi\geq0$ is a constant demonstrating
the coding, decoding, and backhaul power consumption per unit data
rate (W/(bit/s/Hz)) \cite{Emil2015EE}.

The corresponding EE maximization problem under the practical power
consumption model is formulated as 
\begin{align}
\mathop{\max}\limits _{\mathbf{F}\in\mathcal{S}} & \;\;\frac{f\left(\mathbf{F}\right)}{g\left(\mathbf{F}\right)}.\label{RS-EE-problem}
\end{align}

Problem (\ref{RS-EE-problem}) is a fractional programming with non-concave
numerator and non-convex denominator, which is hard to be solved directly.
In addition, the non-convex practical power consumption model with
non-linear rate expression makes the entire problem more complex to
solve.

\subsection{Spectral Efficiency and Energy Efficiency Tradeoff}

The system SE is an increasing function of the transmit power consumption
and its maximization is achieved by consuming all available transmit
power. Such a strategy may not be suitable for EE maximization, since
EE tries to balance SE and power consumption. Hence, EE and SE are
two conflicting metrics in the moderate and high SNR regimes, which
results in a SE-EE tradeoff. In this subsection, we characterize this
tradeoff and identify the precoder strategy that achieves the best
SE-EE tradeoff. The SE-EE tradeoff is an MOO problem, which is given
by 
\begin{align}
\mathop{\max}\limits _{\mathbf{F}\in\mathcal{S}} & \;\;\left[\frac{f\left(\mathbf{F}\right)}{g\left(\mathbf{F}\right)},f\left(\mathbf{F}\right)\right].\label{RS-tradeoff-MOO}
\end{align}

Note that the SE maximization Problem (\ref{RS-SE-problem}) and the
EE maximization Problem (\ref{RS-EE-problem}) are actually special
cases of the SE-EE tradeoff Problem (\ref{RS-tradeoff-MOO}). As a
consequence, individual problems (\ref{RS-SE-problem}) and (\ref{RS-EE-problem})
can also be solved under the framework of our proposed algorithm for
Problem (\ref{RS-tradeoff-MOO}).

In the following, we solve the MOO Problem (\ref{RS-tradeoff-MOO})
by two methods. In particular, we first directly adopt the weighted-sum
method to obtain a weakly Pareto optimal solution\footnote{According to Proposition 3.9 in \cite{book-MO}, the optimal solution
of the corresponding weighted-sum problem is weakly Pareto optimal
for the original MOO problem.}{} of Problem (\ref{RS-tradeoff-MOO}) in Section III. In Section
IV, inspired by the weighted-sum method, a weighted-power method is
developed to characterize the SE-EE tradeoff, which is demonstrated
to have the same performance as the weighted-sum method in the numerical
results section.

\section{Weighed-Sum Approach}

In this section, the weighted-sum approach to solve Problem (\ref{RS-tradeoff-MOO})
is specified. Weighted-sum approach is a classical method to solve
MOO. It converts MOO into a SOO problem by assigning a weight to each
normalized objective function and sum them up. Therefore, the MOO
of (\ref{RS-tradeoff-MOO}) is transformed into 
\begin{align}
\mathop{\max}\limits _{\mathbf{F}\in\mathcal{S}} & \;\;w\frac{f\left(\mathbf{F}\right)}{g\left(\mathbf{F}\right)}+(1-w)\frac{f\left(\mathbf{F}\right)}{\mathrm{P_{c}}},\label{RS-tradeoff-SOO}
\end{align}
where $w\in[0,1]$ is a constant to characterize the priorities of
EE and SE and is decided according to the application environments
and quality of service (QoS) requirements. As specified in (\ref{power}),
$\mathrm{P_{c}}$ is a constant representing the rate-independent
fixed power consumption. According to \cite{Tervo2018tradeoff}, the
denominator $\mathrm{P_{c}}$ of the second fraction in (\ref{RS-tradeoff-SOO})
is introduced to unify the units of the two objective functions in
(\ref{RS-tradeoff-MOO}) so that the values are comparable. We remark
that the denominator of the second term is a constant which could
be chosen randomly without affecting the solutions of Problem (\ref{RS-tradeoff-SOO}).

\subsection{Low-Complexity Algorithm for Two-User System}

In this section, a low-complexity algorithm with closed-form solution
is developed for a two-user system by extending the SE analysis of
\cite{Bruno-WCL} to the SE-EE tradeoff. Based on the proposed low-complexity
algorithm, we gain important insights on the EE-SE trade-off problem.

To that end, the precoding vectors for the private streams are defined
as $\mathbf{f}_{k}=\sqrt{p_{k}}\mathbf{v}_{k},k=1,2$, where $p_{k}$
is the transmit power allocated to the $k$-th private stream, and
$\mathbf{v}_{k}$ is the precoding direction which is fixed by using
ZF \cite{ZF-1} method. Specifically, $\mathbf{v}_{k}=\frac{\hat{\mathbf{v}}_{k}}{||\hat{\mathbf{v}}_{k}||_{2}},k=1,2$,
where $[\hat{\mathbf{v}}_{1},\hat{\mathbf{v}}_{2}]=\mathbf{H}(\mathbf{H}^{\mathrm{H}}\mathbf{H})^{-1}$
and $\mathbf{H}=[\mathbf{h}_{1},\mathbf{h}_{2}]$. Thus, we have $|\mathbf{h}_{2}^{\mathrm{H}}\mathbf{f}_{1}|=0$,
$|\mathbf{h}_{1}^{\mathrm{H}}\mathbf{f}_{2}|=0$, and $|\mathbf{h}_{k}^{\mathrm{H}}\mathbf{f}_{k}|^{2}=\rho p_{k}||\mathbf{h}_{k}||^{2},k=1,2$,
where $\rho=1-|\bar{\mathbf{h}}_{1}^{\mathrm{H}}\bar{\mathbf{h}}_{2}|^{2}$
and $\bar{\mathbf{h}}_{k}=\frac{\mathbf{h}_{k}}{||\mathbf{h}_{k}||_{2}},k=1,2$.
Furthermore, assume that the sum rate of the SE-EE tradeoff problem
in (\ref{RS-tradeoff-SOO}) is maximized by using transmit power $||\mathbf{F}||_{F}^{2}=p_{x}$
(where $p_{x}\leq\mathrm{P_{\mathrm{max}}}$), where the private streams
are allocated with transmit power $p_{1},p_{2}$ such that $p_{1}+p_{2}=tp_{x}$
with $0\leq t\leq1$, and the common stream is assigned with transmit
power $(1-t)p_{x}$. With fixed $t$ and $p_{x}$, the optimal power
allocation for the private streams to maximize the sum rate is the
Water-Filling solution, which is given by\begin{subequations}\label{WF-solution}
\begin{align}
p_{1} & =\max\left(\mu(t,p_{x})-\frac{1}{\beta_{1}},0\right),\label{wf1}\\
p_{2} & =\max\left(\mu(t,p_{x})-\frac{1}{\beta_{2}},0\right),\label{eq:wf2}
\end{align}
\end{subequations} where $\beta_{1}=\frac{\rho||\mathbf{h}_{1}||^{2}}{\sigma_{1}^{2}}$
and $\beta_{2}=\frac{\rho||\mathbf{h}_{2}||^{2}}{\sigma_{2}^{2}}$.
The water level $\mu(t,p_{x})$ is chosen such that $p_{1}+p_{2}=tp_{x}$,
and set as $\mu(t,p_{x})=\frac{tp_{x}}{2}+\frac{1}{2\beta_{1}}+\frac{1}{2\beta_{2}}$.

The precoding vector for the common stream is designed by\begin{subequations}\label{eq:low-pro}
\begin{align}
\mathop{\max}\limits _{\mathbf{f}_{\mathrm{c}}} & \min\left(\frac{|\mathbf{h}_{1}^{\mathrm{H}}\mathbf{f}_{\mathrm{c}}|^{2}}{\sigma_{1}^{2}+|\mathbf{h}_{1}^{\mathrm{H}}\mathbf{f}_{1}|^{2}},\frac{|\mathbf{h}_{2}^{\mathrm{H}}\mathbf{f}_{\mathrm{c}}|^{2}}{\sigma_{2}^{2}+|\mathbf{h}_{2}^{\mathrm{H}}\mathbf{f}_{2}|^{2}}\right)\\
\mathrm{s.t.} & \thinspace\thinspace||\mathbf{f}_{\mathrm{c}}||^{2}=(1-t)p_{x}.
\end{align}
\end{subequations}

According to \cite{Bruno-WCL}, when $p_{1}>0$ and $p_{2}>0$, Problem
(\ref{eq:low-pro}) is equivalent to 
\begin{equation}
\mathop{\max}\limits _{\mathbf{f}_{\mathrm{c}}}\min\left(|\bar{\mathbf{h}}_{1}^{\mathrm{H}}\mathbf{f}_{\mathrm{c}}|^{2},|\bar{\mathbf{h}}_{2}^{\mathrm{H}}\mathbf{f}_{\mathrm{c}}|^{2}\right),\mathrm{s.t.}\thinspace\thinspace||\mathbf{f}_{\mathrm{c}}||^{2}=(1-t)p_{x}.\label{eq:low-pro-1}
\end{equation}
See \cite{Bruno-WCL} for more details about the equivalence of Problems
(\ref{eq:low-pro}) and (\ref{eq:low-pro-1}). Then, it follows from
\cite{two-userL} that the optimal $\mathbf{f}_{\mathrm{c}}$ satisfies
the equations $\frac{|\mathbf{h}_{1}^{\mathrm{H}}\mathbf{f}_{\mathrm{c}}|^{2}}{\sigma_{1}^{2}+|\mathbf{h}_{1}^{\mathrm{H}}\mathbf{f}_{1}|^{2}}=\frac{|\mathbf{h}_{2}^{\mathrm{H}}\mathbf{f}_{\mathrm{c}}|^{2}}{\sigma_{2}^{2}+|\mathbf{h}_{2}^{\mathrm{H}}\mathbf{f}_{2}|^{2}}$
and $|\bar{\mathbf{h}}_{1}^{\mathrm{H}}\mathbf{f}_{\mathrm{c}}|^{2}=|\bar{\mathbf{h}}_{2}^{\mathrm{H}}\mathbf{f}_{\mathrm{c}}|^{2}$,
and the solution of Problem (\ref{eq:low-pro-1}) is $\mathbf{f}_{\mathrm{c}}=\sqrt{(1-t)p_{x}}\mathbf{v}_{\mathrm{c}}$
with precoding direction $\mathbf{v}_{\mathrm{c}}$, 
\begin{equation}
\mathbf{v}_{\mathrm{c}}=\frac{1}{\sqrt{2(1+|\bar{\mathbf{h}}_{1}^{\mathrm{H}}\bar{\mathbf{h}}_{2}|)}}\left(\bar{\mathbf{h}}_{1}+\bar{\mathbf{h}}_{2}e^{-j\angle\bar{\mathbf{h}}_{1}^{\mathrm{H}}\bar{\mathbf{h}}_{2}}\right).\label{cs}
\end{equation}

The sum rate is expressed as $f(t,p_{x})=\log_{2}(1+\beta_{1}p_{1})+\log_{2}(1+\beta_{2}p_{2})+R_{{\rm c}}$,
where $R_{{\rm c}}=\min(\log_{2}(1+\frac{p_{\mathrm{c}}|\mathbf{h}_{1}^{\mathrm{H}}\mathbf{v}_{\mathrm{c}}|^{2}}{\sigma_{1}^{2}+p_{1}|\mathbf{h}_{1}^{\mathrm{H}}\mathbf{v}_{1}|^{2}}),\log_{2}(1+\frac{p_{\mathrm{c}}|\mathbf{h}_{2}^{\mathrm{H}}\mathbf{v}_{\mathrm{c}}|^{2}}{\sigma_{2}^{2}+p_{2}|\mathbf{h}_{2}^{\mathrm{H}}\mathbf{v}_{2}|^{2}}))$.
Owing to the equation $\log_{2}(1+\frac{p_{\mathrm{c}}|\mathbf{h}_{1}^{\mathrm{H}}\mathbf{v}_{\mathrm{c}}|^{2}}{\sigma_{1}^{2}+p_{1}|\mathbf{h}_{1}^{\mathrm{H}}\mathbf{v}_{1}|^{2}})=\log_{2}(1+\frac{p_{\mathrm{c}}|\mathbf{h}_{2}^{\mathrm{H}}\mathbf{v}_{\mathrm{c}}|^{2}}{\sigma_{2}^{2}+p_{2}|\mathbf{h}_{2}^{\mathrm{H}}\mathbf{v}_{2}|^{2}})$,
we can choose $R_{{\rm c}}=\log_{2}(1+\frac{p_{\mathrm{c}}|\mathbf{h}_{2}^{\mathrm{H}}\mathbf{v}_{\mathrm{c}}|^{2}}{\sigma_{2}^{2}+p_{2}|\mathbf{h}_{2}^{\mathrm{H}}\mathbf{v}_{2}|^{2}})$,
and rewrite the sum rate as 
\begin{equation}
f\left(t,p_{x}\right)=\log_{2}\left(1+\beta_{1}p_{1}\right)+\log_{2}\left(1+\beta_{2}p_{2}+\beta_{\mathrm{c}}(1-t)p_{x}\right),\label{eq:rate-2}
\end{equation}
where $\beta_{\mathrm{c}}=\frac{|\mathbf{h}_{2}^{\mathrm{H}}\mathbf{v}_{\mathrm{c}}|^{2}}{\sigma_{2}^{2}}$.

Substituting (\ref{WF-solution}) into (\ref{eq:rate-2}), we can
write 
\begin{equation}
f(t,p_{x})=\log_{2}\left(\beta_{1}\mu(t,p_{x})\right)+\log_{2}\left(\beta_{2}\mu(t,p_{x})+\beta_{\mathrm{c}}(1-t)p_{x}\right).\label{eq:rate-2-1}
\end{equation}
The optimal $t$ that maximizes $f(t,p_{x})$ is obtained by setting
$\frac{\partial f(t,p_{x})}{\partial t}=0$, which results 
\begin{equation}
t=\begin{cases}
\frac{1}{2\beta_{\mathrm{c}}-\beta_{2}}\left(\frac{(\beta_{2}-\beta_{\mathrm{c}})(\beta_{1}+\beta_{2})}{\beta_{1}\beta_{2}p_{x}}+\beta_{\mathrm{c}}\right), & \textrm{if }\frac{1}{2}\leq|\bar{\mathbf{h}}_{1}^{\mathrm{H}}\bar{\mathbf{h}}_{2}|\leq1\\
1, & \textrm{if }0\leq|\bar{\mathbf{h}}_{1}^{\mathrm{H}}\bar{\mathbf{h}}_{2}|<\frac{1}{2},
\end{cases}\label{eq:t}
\end{equation}
which is due to $t-1=\frac{\beta_{2}-\beta_{\mathrm{c}}}{2\beta_{\mathrm{c}}-\beta_{2}}(\frac{\beta_{1}+\beta_{2}}{\beta_{1}\beta_{2}p_{x}}+1),$
$2\beta_{\mathrm{c}}-\beta_{2}>0$, and 
\begin{align*}
(\beta_{2}-\beta_{\mathrm{c}} & )\begin{cases}
\leq0, & \textrm{if }\frac{1}{2}\leq|\bar{\mathbf{h}}_{1}^{\mathrm{H}}\bar{\mathbf{h}}_{2}|\leq1\\
>0, & \textrm{if }0\leq|\bar{\mathbf{h}}_{1}^{\mathrm{H}}\bar{\mathbf{h}}_{2}|<\frac{1}{2}.
\end{cases}
\end{align*}
The solution in (\ref{eq:t}) maximizing SE is the same with the solution
in (13) in \cite{Bruno-WCL}. According to (\ref{eq:t}), when $0\leq|\bar{\mathbf{h}}_{1}^{\mathrm{H}}\bar{\mathbf{h}}_{2}|<\frac{1}{2}$,
RSMA boils down to SDMA ($t=1$).

Substituting (\ref{eq:t}) into (\ref{WF-solution}), then the transmit
power of each stream is \begin{subequations}\label{WF-solution-1}
\begin{align}
p_{1} & =\max\left(\frac{\beta_{\mathrm{c}}}{2\left(2\beta_{\mathrm{c}}-\beta_{2}\right)}\left(\frac{2\beta_{2}}{\beta_{1}\beta_{\mathrm{c}}}+\frac{1}{\beta_{2}}-\frac{3}{\beta_{1}}+p_{x}\right),0\right),\label{wf1-1}\\
p_{2} & =\max\left(\frac{\beta_{\mathrm{c}}}{2\left(2\beta_{\mathrm{c}}-\beta_{2}\right)}\left(\frac{2}{\beta_{\mathrm{c}}}+\frac{1}{\beta_{1}}-\frac{3}{\beta_{2}}+p_{x}\right),0\right).\label{eq:wf2-1}
\end{align}
\end{subequations}

Furthermore, substituting (\ref{WF-solution-1}) into (\ref{eq:rate-2}),
the sum rate is expressed as 
\begin{align}
f(p_{x})= & \log_{2}\left(1+\beta_{1}p_{1}\right)\nonumber \\
 & +\log_{2}\left(1+\beta_{2}p_{2}+\beta_{\mathrm{c}}(p_{x}-p_{1}-p_{2})\right)\nonumber \\
= & \log_{2}\left(\frac{\beta_{\mathrm{c}}\beta_{1}}{2\left(2\beta_{\mathrm{c}}-\beta_{2}\right)}\left(p_{x}+\frac{1}{\beta_{1}}+\frac{1}{\beta_{2}}\right)\right)\nonumber \\
 & +\log_{2}\left(\frac{\beta_{\mathrm{c}}}{2}\left(p_{x}+\frac{1}{\beta_{1}}+\frac{1}{\beta_{2}}\right)\right).\label{eq:rate-px}
\end{align}

With (\ref{eq:rate-px}), the SE-EE tradeoff problem in (\ref{RS-tradeoff-SOO})
is reformulated as\begin{subequations}\label{eq:final-p} 
\begin{align}
\mathop{\max}\limits _{p_{x}} & \;\;L(p_{x})=w\frac{f(p_{x})}{p_{x}+\chi f(p_{x})+\mathrm{P_{c}}}+(1-w)\frac{f(p_{x})}{\mathrm{P_{c}}}\label{eq:ddd}\\
\textrm{s.t.} & \;\;0\leq p_{x}\leq\mathrm{P_{\mathrm{max}}}.
\end{align}
\end{subequations}Problem (\ref{eq:final-p}) is a convex problem
with the concave objective function $L(p_{x})$ and the affine constraint.
The first-order derivative of $L(p_{x})$ is a monotone decreasing
function of $p_{x}$, and is expressed as 
\[
L^{'}(p_{x})=\frac{f^{'}(p_{x})}{\mathrm{P_{c}}}\left(1-\left(\frac{z_{1}(p_{x})}{z_{2}(p_{x})}+1\right)w\right),
\]
where $f^{'}(p_{x})=\frac{2}{\ln2(p_{x}+\frac{1}{\beta_{1}}+\frac{1}{\beta_{2}})}>0$,
$z_{1}(p_{x})=\mathrm{P_{c}}(f(p_{x})-f^{'}(p_{x})p_{x}-f^{'}(p_{x})\mathrm{P_{c}})$,
and $z_{2}(p_{x})=f^{'}(p_{x})(p_{x}+\chi f(p_{x})+\mathrm{P_{c}})^{2}>0$.
When $w=1$, we denote $\widetilde{p}_{w=1}$ which meets $L^{'}(\widetilde{p}_{w=1})=0$
and $z_{1}(\widetilde{p}_{w=1})=0$. Since $-z_{1}^{'}(p_{x})=\mathrm{P_{c}}f^{''}(p_{x})(p_{x}+\mathrm{P_{c}})<0$
with $f^{''}(p_{x})=\frac{-2}{(\ln2(p_{x}+\frac{1}{\beta_{1}}+\frac{1}{\beta_{2}}))^{2}}<0$,
then $z_{1}(p_{x})$ is a monotone decreasing function of $p_{x}$.
The minimum value of $L^{'}(p_{x})$ is obtained at $p_{x}=\mathrm{P_{\mathrm{max}}}$.
If $\mathrm{P_{\mathrm{max}}}\leq\widetilde{p}_{w=1}$, we always
have $L^{'}(\mathrm{P_{\mathrm{max}}})\geq0$, then the optimal $p_{x}$
is $\mathrm{P_{\mathrm{max}}}$. If $\mathrm{P_{\mathrm{max}}}>\widetilde{p}_{w=1}$,
we have $L^{'}(\mathrm{P_{\mathrm{max}}})>0$ when $0\leq w<\frac{z_{2}(\mathrm{P_{\mathrm{max}}})}{z_{1}(\mathrm{P_{\mathrm{max}}})+z_{2}(\mathrm{P_{\mathrm{max}}})}$,
then the optimal $p_{x}$ is $\mathrm{P_{\mathrm{max}}}$. If $\mathrm{P_{\mathrm{max}}}>\widetilde{p}_{w=1}$,
we have $L^{'}(\mathrm{P_{\mathrm{max}}})\leq0$ when $\frac{z_{2}(\mathrm{P_{\mathrm{max}}})}{z_{1}(\mathrm{P_{\mathrm{max}}})+z_{2}(\mathrm{P_{\mathrm{max}}})}\leq w\leq1$,
then the optimal $p_{x}$ is the solution of $L^{'}(p_{x})=0$. Denote
$\widetilde{p}$ such that $L^{'}(\widetilde{p})=0$ when $\mathrm{P_{\mathrm{max}}}>\widetilde{p}_{w=1}$.
$\widetilde{p}_{w=1}$ and $\widetilde{p}$ can be obtained by ``solve''
function in MATLAB. Please note that $\widetilde{p}_{w=1}$ and $\widetilde{p}$
is globle optimal due to the concave function $L(p_{x})$. Based on
the above analysis, the optimal solution of Problem (\ref{eq:final-p})
is given by 
\begin{equation}
\mathrm{P_{\mathrm{opt}}}=\begin{cases}
\mathrm{P_{\mathrm{max}}}, & \textrm{if }\mathrm{P_{\mathrm{max}}}\leq\widetilde{p}_{w=1}\\
\mathrm{P_{\mathrm{max}}}, & \textrm{if }\mathrm{P_{\mathrm{max}}}>\widetilde{p}_{w=1}\textrm{ and }0\leq w<\frac{1}{\frac{z_{1}(\mathrm{P_{\mathrm{max}}})}{z_{2}(\mathrm{P_{\mathrm{max}}})}+1}\\
\widetilde{p}, & \textrm{if }\mathrm{P_{\mathrm{max}}}>\widetilde{p}_{w=1}\textrm{ and }\frac{1}{\frac{z_{1}(\mathrm{P_{\mathrm{max}}})}{z_{2}(\mathrm{P_{\mathrm{max}}})}+1}\leq w\leq1.
\end{cases}\label{eq:opt-power}
\end{equation}
The corresponding optimal sum rate and the energy efficiency are directly
given by $\textrm{SE}=f(\mathrm{P_{\mathrm{opt}}})$ and $\textrm{EE}=\frac{f(\mathrm{P_{\mathrm{opt}}})}{\mathrm{P_{\mathrm{opt}}}+\chi f(\mathrm{P_{\mathrm{opt}}})+\mathrm{P_{c}}}$,
respectively.

\textbf{Remark 2}: Following from (\ref{eq:opt-power}), we would
like to highlight that the optimal transmit power $\mathrm{P_{\mathrm{opt}}}$
of the SE-EE tradeoff is controlled by two factors, namely, the maximum
transmit power limit $\mathrm{P_{\mathrm{max}}}$ at the BS and the
weight $w$ allocated to EE and SE. When $\mathrm{P_{\mathrm{max}}}\leq\widetilde{p}_{w=1}$,
SE and EE are both monotonic increasing function of the transmit power
for all weights in $0\leq w\leq1$. In contrast, when $\mathrm{P_{\mathrm{max}}}>\widetilde{p}_{w=1}$,
EE and SE are conflicting with each other as the transmit power increases.
Therefore, such result helps us further show the necessity of investigating
the SE-EE tradeoff problem when $\mathrm{P_{\mathrm{max}}}>\widetilde{p}_{w=1}$,
and the optimal $\mathrm{P_{\mathrm{opt}}}$ that achieves the optimal
SE-EE tradeoff is adjusted by $w$.

\subsection{SCA-based Algorithm for $K$-user system}

Problem (\ref{RS-tradeoff-SOO}) for the $K$-user system is difficult
to solve for the reason that the objective function has non-convex
numerator and denominator, and the non-convex property mainly comes
from the rate expressions. A popular method to solve general fractional
programming problem is Dinkelbach's method \cite{FP-Dinkelbach}.
It is capable of converting fractional programming to linear programming
by introducing parameters to denote those fractions. These parametric
linear programming problems are then addressed by applying SCA \cite{Yang2019}
and WMMSE \cite{Shiwen2013letter}. The Dinkelbach's framework is
in fact a two-layer iterative procedure, which requires huge computational
complexity. More importantly, the convergence of this framework cannot
be guaranteed since each parametric linear programming problem may
only achieve local optimum \cite{Oskari2018}.

In order to address the above shortcomings, we propose a one-layer
iterative algorithm under the SCA framework \cite{Razaviyayn2014PHD}
directly. Firstly, we introduce a new variable $\eta$ denoting the
EE and reformulate Problem (\ref{RS-tradeoff-SOO}) as \begin{subequations}\label{RS-tradeoff-SOO-problem}
\begin{align}
\mathop{\max}\limits _{\mathbf{F}\in\mathcal{S},\eta} & \;\;w\eta+(1-w)\frac{f\left(\mathbf{F}\right)}{\mathrm{P_{c}}}\label{RS-tradeoff-SOO-1}\\
\textrm{s.t.}\ \ \  & \eta\leq\frac{f\left(\mathbf{F}\right)}{g\left(\mathbf{F}\right)}.\label{RS-tradeoff-SOO-constrain}
\end{align}
\end{subequations}

To tractably recast non-convex fractional constraint (\ref{RS-tradeoff-SOO-constrain}),
we replace it with the following three constraints \begin{subequations}
\begin{align}
\eta & \leq\frac{x^{2}}{y},\label{eq:RS-tradeoff-SOO-constrain-1}\\
x^{2} & \leq f\left(\mathbf{F}\right),\label{eq:RS-tradeoff-SOO-constrain-2}\\
g\left(\mathbf{F}\right) & \leq y,\label{eq:RS-tradeoff-SOO-constrain-3}
\end{align}
\end{subequations}where variable $x$ represents the square roof
of the total sum rate, and variable $y$ represents the total power.
Then, we introduce variables $\mathbf{r}=[r_{\mathrm{c}},r_{1},...,r_{K}]^{\mathrm{T}}$
to denote rates $\{R_{i}\left(\mathbf{F}\right)\}_{\forall i\in\mathcal{K}_{\mathrm{c}}}$
and rewrite Problem (\ref{RS-tradeoff-SOO-problem}) equivalently
as \begin{subequations}\label{RS-tradeoff-SOO-problem-1} 
\begin{align}
\mathop{\max}\limits _{\mathbf{f}\in\mathcal{S},\eta,\mathbf{r},x,y} & \;\;w\eta+\frac{(1-w)}{\mathrm{P_{c}}}\sum_{i\in\mathcal{K}_{\mathrm{c}}}r_{i}\label{RS-tradeoff-SOO-1-1}\\
\textrm{s.t.}\ \ \  & \textrm{(\ref{eq:RS-tradeoff-SOO-constrain-1})}\\
 & x^{2}\leq\sum_{i\in\mathcal{K}_{\mathrm{c}}}r_{i}\label{eq:tradeoff-1-c-1}\\
 & ||\mathbf{F}||_{F}^{2}+\mathrm{P_{c}}+\chi\sum_{i\in\mathcal{K}_{\mathrm{c}}}r_{i}\leq y\label{eq:tradeoff-1-c-2}\\
 & r_{k}\leq R_{k}\left(\mathbf{F}\right),\forall k\in\mathcal{K}\thinspace\label{eq:tradeoff-1-c-3}\\
 & r_{\mathrm{c}}\leq R_{\mathrm{c},k}\left(\mathbf{F}\right),\forall k\in\mathcal{K}\label{eq:tradeoff-1-c-4}
\end{align}
\end{subequations}The non-convexity of Problem (\ref{RS-tradeoff-SOO-1-1})
is due to the constraints (\ref{eq:RS-tradeoff-SOO-constrain-1}),
(\ref{eq:tradeoff-1-c-3}) and (\ref{eq:tradeoff-1-c-4}), which motivates
us to use SCA framework to approximate the non-convex constraints.
Specifically the right hand side of (\ref{eq:RS-tradeoff-SOO-constrain-1})
is a quadratic-over-linear function, which is jointly convex in $(x,y)$.
We approximate it by its first-order lower approximation at fixed
point $(x^{(n)},y^{(n)})$ as \cite{book-convex} 
\begin{equation}
\frac{x^{2}}{y}\geq\frac{2x^{(n)}}{y^{(n)}}x-\left(\frac{x^{(n)}}{y^{(n)}}\right)^{2}y\triangleq\phi^{(n)}(x,y).\label{eq:approx-1}
\end{equation}

The remaining challenge is to tackle the non-convexity of constraints
(\ref{eq:tradeoff-1-c-3}) and (\ref{eq:tradeoff-1-c-4}). In most
literature, the relation between rate and WMMSE is used to transform
the non-convex rate-based function into its convex WMMSE counterpart
by introducing auxiliary variables, i.e., weights and equalizers.
The method known as WMMSE is widely used in the literature \cite{Shiwen2013letter,Razaviyayn2014PHD,Joudeh2016Tcom}
and shows good performance. However, WMMSE method is an iterative
optimization method, in which the weights, equalizers and precoders
are updated in an iterative manner.

In the following, we investigate the intrinsic convexity of the rate
expressions $R_{k}\left(\mathbf{F}\right)$ and $R_{\mathrm{c},k}\left(\mathbf{F}\right)$,
and then propose two lower bounds to approximate the non-convex rate
expressions by using the SCA method.

\subsubsection{Lower-bound (LB) I}

The first lower bound of rate is summarized in the following Lemma
\ref{First-surroagte-function-lemma}.

\begin{lemma}\label{First-surroagte-function-lemma} Let $\mathbf{F}^{(n)}$
denote the optimal solution obtained in the $(n-1)$-th iteration.
The concave lower bound function of $R_{k}\left(\mathbf{F}\right)$
in the $n$-th iteration is given by 
\begin{align}
 & R_{k}^{(n)}\left(\mathbf{F}\right)\triangleq\text{\ensuremath{\mathrm{cons}\mathrm{t}_{k}}}+2\text{Re}\left\{ a_{k}\mathbf{b}_{k}^{\mathrm{H}}\mathbf{f}_{k}\right\} -a_{k}\sum_{i\in\mathcal{K}}\mathbf{f}_{i}^{\mathrm{H}}\mathbf{b}_{k}\mathbf{b}_{k}^{\mathrm{H}}\mathbf{f}_{i},\label{eq:approx-2}
\end{align}
at point $\mathbf{F}^{(n)}$, where \begin{subequations} 
\begin{align}
 & a_{k}=1+(\sigma_{k}^{2}+\sum_{i\in\mathcal{K}\backslash\{k\}}\mathbf{h}_{k}^{\mathrm{H}}\mathbf{f}_{i}^{(n)}\mathbf{f}_{i}^{(n),\mathrm{H}}\mathbf{h}_{k})^{-1}\mathbf{f}_{k}^{(n),\mathrm{H}}\mathbf{h}_{k}\mathbf{h}_{k}^{\mathrm{H}}\mathbf{f}_{k}^{(n)},\label{eq:ak}\\
 & \mathbf{b}_{k}=(\sigma_{k}^{2}+\sum_{i\in\mathcal{K}}\mathbf{h}_{k}^{\mathrm{H}}\mathbf{f}_{i}^{(n)}\mathbf{f}_{i}^{(n),\mathrm{H}}\mathbf{h}_{k})^{-1}\mathbf{h}_{k}^{\mathrm{H}}\mathbf{f}_{k}^{(n)}\mathbf{h}_{k},\label{eq:bk}\\
\text{} & \mathrm{cons}\mathrm{t}_{k}=R_{k}\left(\mathbf{F}^{(n)}\right)-2\text{Re}\left\{ a_{k}\mathbf{b}_{k}^{\mathrm{H}}\mathbf{f}_{k}^{(n)}\right\} \nonumber \\
 & \thinspace\thinspace\thinspace\thinspace\thinspace\thinspace\thinspace\thinspace\thinspace\thinspace\thinspace\thinspace\thinspace\thinspace\thinspace\thinspace\thinspace\thinspace\thinspace\thinspace\thinspace\thinspace\thinspace\thinspace\thinspace\thinspace\thinspace\thinspace\thinspace\thinspace\thinspace\thinspace\thinspace\thinspace\thinspace+a_{k}\sum_{i\in\mathcal{K}}\mathbf{f}_{i}^{(n),\mathrm{H}}\mathbf{b}_{k}\mathbf{b}_{k}^{\mathrm{H}}\mathbf{f}_{i}^{(n)}.\label{eq:const-MM}
\end{align}
\end{subequations}

Meanwhile, the concave lower bound function of $R_{\mathrm{c},k}\left(\mathbf{F}\right)$
is given by 
\begin{align}
 & R_{\mathrm{c},k}^{(n)}\left(\mathbf{F}\right)\triangleq\nonumber \\
 & \mathrm{cons}\mathrm{t}_{\mathrm{c},k}+2\text{Re}\left\{ a_{\mathrm{c},k}\mathbf{b}_{\mathrm{c},k}^{\mathrm{H}}\mathbf{f}_{\mathrm{c}}\right\} -a_{\mathrm{c},k}\sum_{i\in\mathcal{K}_{\mathrm{c}}}\mathbf{f}_{i}^{\mathrm{H}}\mathbf{b}_{\mathrm{c},k}\mathbf{b}_{\mathrm{c},k}^{\mathrm{H}}\mathbf{f}_{i},\label{eq:approx-3}
\end{align}
at point $\mathbf{f}^{n}$, where \begin{subequations} 
\begin{align}
 & a_{\mathrm{c},k}=1+(\sigma_{k}^{2}+\sum_{i\in\mathcal{K}}\mathbf{h}_{k}^{\mathrm{H}}\mathbf{f}_{i}^{(n)}\mathbf{f}_{i}^{(n),\mathrm{H}}\mathbf{h}_{k})^{-1}\mathbf{f}_{\mathrm{c}}^{(n),\mathrm{H}}\mathbf{h}_{k}\mathbf{h}_{k}^{\mathrm{H}}\mathbf{f}_{\mathrm{c}}^{(n)},\\
 & \mathbf{b}_{\mathrm{c},k}=(\sigma_{k}^{2}+\sum_{i\in\mathcal{K}_{\mathrm{c}}}\mathbf{h}_{k}^{\mathrm{H}}\mathbf{f}_{i}^{(n)}\mathbf{f}_{i}^{(n),\mathrm{H}}\mathbf{h}_{k})^{-1}\mathbf{h}_{k}^{\mathrm{H}}\mathbf{f}_{\mathrm{c}}^{(n)}\mathbf{h}_{k},\\
\text{} & \mathrm{cons}\mathrm{t}_{\mathrm{c},k}=R_{\mathrm{c},k}\left(\mathbf{F}^{(n)}\right)-2\text{Re}\left\{ a_{\mathrm{c},k}\mathbf{b}_{\mathrm{c},k}^{\mathrm{H}}\mathbf{f}_{\mathrm{c}}^{(n)}\right\} \nonumber \\
 & \ \ \thinspace\thinspace\thinspace\thinspace\thinspace\thinspace\thinspace\thinspace\thinspace\thinspace\thinspace\thinspace\thinspace\thinspace\thinspace\thinspace\thinspace\thinspace\thinspace\thinspace\thinspace\thinspace\thinspace\thinspace\thinspace\thinspace\thinspace\thinspace\thinspace\thinspace\thinspace\thinspace\thinspace\thinspace\thinspace\thinspace\thinspace\thinspace\thinspace+a_{\mathrm{c},k}\sum_{i\in\mathcal{K}_{\mathrm{c}}}\mathbf{f}_{i}^{(n),\mathrm{H}}\mathbf{b}_{\mathrm{c},k}\mathbf{b}_{\mathrm{c},k}^{\mathrm{H}}\mathbf{f}_{i}^{(n)}.
\end{align}
\end{subequations} \end{lemma}

\textbf{\textit{Proof: }}Please refer to Appendix \ref{Proof-lemma-2}.\hspace{3cm}$\blacksquare$

With the concave lower bound approximations (\ref{eq:approx-1}),
(\ref{eq:approx-3}), and (\ref{eq:approx-2}), Problem (\ref{RS-tradeoff-SOO-problem-1})
is reformulated equivalently as the following convex problem: \begin{subequations}\label{RS-tradeoff-SOO-problem-2}
\begin{align}
\mathop{\max}\limits _{\mathbf{F}\in\mathcal{S},\eta,\mathbf{r},x,y} & \;\;w\eta+\frac{(1-w)}{\mathrm{P_{c}}}\sum_{i\in\mathcal{K}_{\mathrm{c}}}r_{i}\label{RS-tradeoff-SOO-1-2}\\
\textrm{s.t.}\ \ \  & \eta\leq\phi^{(n)}(x,y)\label{eq:const-2-1}\\
 & r_{k}\leq R_{k}^{(n)}\left(\mathbf{F}\right),\forall k\in\mathcal{K}\label{eq:const-2-2}\\
 & r_{\mathrm{c}}\leq R_{\mathrm{c},k}^{(n)}\left(\mathbf{F}\right),\forall k\in\mathcal{K}\label{eq:const-2-3}\\
 & \textrm{(\ref{eq:tradeoff-1-c-1}),(\ref{eq:tradeoff-1-c-2})}.
\end{align}
\end{subequations}

\subsubsection{Lower-bound (LB) II}

We note that the lower bound approximation proposed in Lemma \ref{First-surroagte-function-lemma}
is the first-order approximation of log function directly. Recall
that each rate function is a composition function with an inner fractional
SINR function embraced by an outer log function. Motivated by the
convexity of the outer log function, another method is to approximate
only the inner SINR function with its concave lower bound and keep
the outer log function. Specifically, we introduce new variables $\boldsymbol{\gamma}=[\gamma_{1},...,\gamma_{K}]^{\mathrm{T}}$
to denote the SINRs of the private streams at all users and variables
$\boldsymbol{\gamma}_{\mathrm{c}}=[\gamma_{\mathrm{c},1},...,\gamma_{\mathrm{c},K}]^{\mathrm{T}}$
to denote the SINRs of the common streams at all users. Constraints
(\ref{eq:tradeoff-1-c-3}) and (\ref{eq:tradeoff-1-c-4}) can be recast
equivalently as \begin{subequations} 
\begin{align}
 & r_{k}\leq\log_{2}(1+\gamma_{k}),\forall k\in\mathcal{K}\thinspace\label{eq:low-1}\\
 & r_{\mathrm{c}}\leq\log_{2}(1+\gamma_{\mathrm{c},k}),\forall k\in\mathcal{K}\label{eq:low-2}\\
 & \gamma_{k}\leq\frac{\left|\mathbf{h}_{k}^{\mathrm{H}}\mathbf{f}_{k}\right|^{2}}{I_{-k}},\forall k\in\mathcal{K}\thinspace\label{eq:low-3}\\
 & \gamma_{\mathrm{c},k}\leq\frac{\left|\mathbf{h}_{k}^{\mathrm{H}}\mathbf{f}_{\mathrm{c}}\right|^{2}}{I_{k}},\forall k\in\mathcal{K}\label{eq:low-4}
\end{align}
\end{subequations}where $I_{-k}=\sigma_{k}^{2}+\sum_{i\in\mathcal{K}\backslash\{k\}}\mathbf{h}_{k}^{\mathrm{H}}\mathbf{f}_{i}\mathbf{f}_{i}^{\mathrm{H}}\mathbf{h}_{k}$
and $I_{k}=I_{-k}+\mathbf{h}_{k}^{\mathrm{H}}\mathbf{f}_{k}\mathbf{f}_{k}^{\mathrm{H}}\mathbf{h}_{k}$.
The right hand side of (\ref{eq:low-3}) and (\ref{eq:low-4}) are
all in the form of $\frac{x^{2}}{y}$. Thus, according to (\ref{eq:approx-1}),
we obtain (\ref{eq:low-3-1}) by applying substitutions $x=\mathbf{h}_{k}^{\mathrm{H}}\mathbf{f}_{k}$
and $y=I_{-k}$, and (\ref{eq:low-4-1}) by applying substitutions
$x=\mathbf{h}_{k}^{\mathrm{H}}\mathbf{f}_{\mathrm{c}}$ and $y=I_{k}$.
\begin{align}
\frac{\left|\mathbf{h}_{k}^{\mathrm{H}}\mathbf{f}_{k}\right|^{2}}{I-k} & \geq\frac{2\textrm{Re}\left\{ \mathbf{f}_{k}^{n,\mathrm{H}}\mathbf{h}_{k}\mathbf{h}_{k}^{\mathrm{H}}\mathbf{f}_{k}\right\} }{I_{-k}^{n}}-\left|\frac{\mathbf{h}_{k}^{\mathrm{H}}\mathbf{f}_{k}^{n}}{I_{-k}^{n}}\right|^{2}I_{-k}\triangleq\varGamma_{k}^{(n)}(\mathbf{F}),\label{eq:low-3-1}\\
\frac{\left|\mathbf{h}_{k}^{\mathrm{H}}\mathbf{f}_{\mathrm{c}}\right|^{2}}{I_{k}} & \geq\frac{2\textrm{Re}\left\{ \mathbf{f}_{\mathrm{c}}^{n,\mathrm{H}}\mathbf{h}_{k}\mathbf{h}_{k}^{\mathrm{H}}\mathbf{f}_{\mathrm{c}}\right\} }{I_{k}^{n}}-\left|\frac{\mathbf{h}_{k}^{\mathrm{H}}\mathbf{f}_{k}^{n}}{I_{k}^{n}}\right|^{2}I_{k}\triangleq\varGamma_{\mathrm{c},k}^{(n)}(\mathbf{F}).\label{eq:low-4-1}
\end{align}

With (\ref{eq:low-1}), (\ref{eq:low-2}), and the concave lower bound
approximations (\ref{eq:low-3-1}) and (\ref{eq:low-4-1}), Problem
(\ref{RS-tradeoff-SOO-1-1}) is reformulated equivalently as the following
convex problem: \begin{subequations}\label{RS-tradeoff-SOO-problem-3}
\begin{align}
\mathop{\max}\limits _{\mathbf{F}\in\mathcal{S},\eta,\mathbf{r},x,y,\boldsymbol{\gamma},\boldsymbol{\gamma}_{\mathrm{c}}} & \;\;w\eta+\frac{(1-w)}{\mathrm{P_{c}}}\sum_{i\in\mathcal{K}_{\mathrm{c}}}r_{i}\label{RS-tradeoff-SOO-1-3}\\
\textrm{s.t.}\ \ \  & \gamma_{k}\leq\varGamma_{k}^{(n)}(\mathbf{F}),\forall k\in\mathcal{K}\label{eq:low-5}\\
 & \gamma_{\mathrm{c},k}\leq\varGamma_{\mathrm{c},k}^{(n)}(\mathbf{F}),\forall k\in\mathcal{K}\label{eq:low-6}\\
 & (\ref{eq:tradeoff-1-c-1}),(\ref{eq:tradeoff-1-c-2}),(\ref{eq:const-2-1}),(\ref{eq:low-1}),(\ref{eq:low-2}).
\end{align}
\end{subequations}

LB II approximating only the inner SINR function is tighter than LB
I which is the lower bound of the outer rate function. Consequently,
LB II requires less number of iterations for convergence. In the following,
we derive algorithms based on the specified LB I or LB II, and the
relation between LB I and LB II can be reflected clearly in the simulation.
The optimal SE-EE tradeoff problem can be achieved by designing $\mathbf{f}$
under the SCA framework which is summarized in Algorithm \ref{algorithm-SCA}.
The convex Problem (\ref{RS-tradeoff-SOO-problem-2}) and (\ref{RS-tradeoff-SOO-problem-3})
are solved by the standard interior-point algorithm \cite{book-convex}.
The initial points are generated as follows. $\mathbf{F}^{(0)}$ is
created to meet the power constraint $\mathbf{F}\in\mathcal{S}$,
and then $x^{(0)}$ and $y^{(0)}$ are obtained by setting constraints
(\ref{eq:RS-tradeoff-SOO-constrain-2}) and (\ref{eq:RS-tradeoff-SOO-constrain-3})
to be equality, respectively.

\begin{algorithm}
\caption{SCA-based precoder design for the Problem (\ref{RS-tradeoff-SOO})}

\label{algorithm-SCA}

\begin{algorithmic}[1]

\REQUIRE Set $n=0$, and generate initialized points ($\mathbf{F}^{(n)},x^{(n)},y^{(n)}$).

\REPEAT

\STATE Update ($\mathbf{F}^{(n+1)},x^{(n+1)},y^{(n+1)}$) according
to (\ref{RS-tradeoff-SOO-problem-2}) or (\ref{RS-tradeoff-SOO-problem-3}).

\STATE $n=n+1$.

\UNTIL Convergence.

\end{algorithmic} 
\end{algorithm}

\subsubsection{Convergence and Complexity Analysis}

In this subsection, we discuss the convergence and per-iteration complexity
of Algorithm \ref{algorithm-SCA}. To start with, the optimal solution
of Problem (\ref{RS-tradeoff-SOO-problem-2}) or (\ref{RS-tradeoff-SOO-problem-3})
obtained at the $n$-th iteration is also feasible at the next iteration
\cite{Marks1978}. Therefore, the sequence of the objective values
generated by Algorithm \ref{algorithm-SCA} is non-decreasing and
the sequence is bounded above due to the power constraints $\mathbf{F}\in\mathcal{S}$.
Hence, the convergence of Algorithm \ref{algorithm-SCA} is guaranteed.
Moreover, the following proposition shows the solution property of
Algorithm \ref{algorithm-SCA}. \begin{proposition}\label{Theorem-socp-kkt}
Denote by $\mathbf{F}^{o}$ the converged solution of Algorithm \ref{algorithm-SCA},
then $\mathbf{F}^{o}$ converges to a KKT point of Problem (\ref{RS-tradeoff-SOO}).
\end{proposition}

\textbf{\textit{Proof: }}Please refer to Appendix \ref{subsec:The-proof-of-5}.\hspace{3cm}$\blacksquare$

Next, we estimate the worst-case per-iteration complexity of Algorithm
1 with the second-order cone programming (SOCP) (\ref{RS-tradeoff-SOO-problem-2})
and general convex program (GCP) (\ref{RS-tradeoff-SOO-problem-3}).
Specifically, the computational complexity of solving SOCP is $\mathcal{O}(N_{\mathrm{socp}}M_{\mathrm{socp}}^{3.5}+N_{\mathrm{socp}}^{3}M_{\mathrm{socp}}^{2.5})$,
where $N_{\mathrm{socp}}$ and $M_{\mathrm{socp}}$ are the dimension
of second order cone and the number of second order cone constraints,
respectively. Therefore, the per-iteration computational complexity
of solving SOCP (\ref{RS-tradeoff-SOO-problem-2}) is $\mathcal{O}(N_{t}(K+1)^{4.5}+N_{t}^{3}(K+1)^{5.5})$.
Meanwhile, the per-iteration computational complexity of solving the
GCP (\ref{RS-tradeoff-SOO-problem-3}) is $\mathcal{O}(N_{t}^{4}(K+1)^{4})$.

\section{Weighted-Power Approach}

In the low transmit power constraint $P_{\mathrm{max}}$, both EE
and SE increase with $P_{\mathrm{max}}$, and the optimal transmit
power equals $P_{\mathrm{max}}$. Therefore, there is no conflict
of interest between these two objective functions. However, when $P_{\mathrm{max}}$
is large, the maximum SE is achieved when the optimal transmit power
equals $P_{\mathrm{max}}$, which reduces the EE. In order to maximize
EE, only part of the available power is used, which reduces the SE
eventually. That means there is a tradeoff between EE and SE moderate
and high SNR regimes. Since both EE and SE are affected by transmit
power, a weighted-power EE metric is proposed to investigate this
tradeoff.

According to \cite{tradeoff2016tvt}, maximizing EE and SE is also
equivalent to minimizing their inverse. Therefore Problem (\ref{RS-tradeoff-MOO})
is equivalent to 
\begin{align}
\mathop{\min}\limits _{\mathbf{F}\in\mathcal{S}} & \;\;\left[\frac{g_{\mathrm{RS}}\left(\mathbf{F}\right)}{f_{\mathrm{RS}}\left(\mathbf{F}\right)},\frac{1}{f_{\mathrm{RS}}\left(\mathbf{F}\right)}\right].\label{RS-tradeoff-MOO-2}
\end{align}
Problem (\ref{RS-tradeoff-MOO-2}) is also solved via its corresponding
single-objective problem as follows 
\begin{align}
\mathop{\min}\limits _{\mathbf{F}\in\mathcal{S}} & \;\;w\frac{g_{\mathrm{RS}}\left(\mathbf{F}\right)}{f_{\mathrm{RS}}\left(\mathbf{F}\right)}+(1-w)\frac{\mathrm{P_{c}}}{f_{\mathrm{RS}}\left(\mathbf{F}\right)},\label{RS-tradeoff-SOO-2}
\end{align}
where $w\in[0,1]$. With the same denominator, Problem (\ref{RS-tradeoff-SOO-2})
is equivalent to 
\begin{align}
\mathop{\max}\limits _{\mathbf{F}\in\mathcal{S}} & \;\;\frac{f_{\mathrm{RS}}\left(\mathbf{F}\right)}{w(||\mathbf{F}||_{F}^{2}+\chi f_{\mathrm{RS}}\left(\mathbf{F}\right))+\mathrm{P_{c}}}.\label{RS-tradeoff-weighted-power-problem}
\end{align}

From a mathematical point of view, the metric in Problem (\ref{RS-tradeoff-weighted-power-problem})
only has an additional constant $w$ in the denominator compared with
Problem (\ref{RS-EE-problem}). Hence, we name the objective of Problem
(\ref{RS-tradeoff-weighted-power-problem}) weighted-power EE metric.
Physically speaking, by changing $w$ from 0 to 1, we could investigate
the SE-EE tradeoff. When $w=0$, Problem (\ref{RS-tradeoff-weighted-power-problem})
focuses only on maximizing the SE without considering how much energy
is consumed. When $w$ increases, it means there is a penalty for
increasing the SNR. This process may reduce the power consumption
and thereby reduce the SE. When $w=1$, Problem (\ref{RS-tradeoff-weighted-power-problem})
focuses only on maximizing the EE.

When $K=2$, a closed-form solution of Problem (\ref{RS-tradeoff-weighted-power-problem})
can be obtained by setting $L(p_{x})$ in Problem (\ref{eq:final-p})
as 
\begin{equation}
L_{2}(p_{x})=\frac{f(p_{x})}{w(p_{x}+\chi f(p_{x}))+\mathrm{P_{c}}}.\label{eq:power-l}
\end{equation}
(\ref{eq:power-l}) is concave and its first-order derivative is a
monotone decreasing function of $p_{x}$. We have
\begin{align}
L_{[2]}^{'}(p_{x}) & =\frac{z_{3}(p_{x})w+f^{'}(p_{x})\mathrm{P_{c}}}{\left(w(p_{x}+\chi f(p_{x}))+\mathrm{P_{c}}\right)^{2}},
\end{align}
where $z_{3}(p_{x})=f^{'}(p_{x})p_{x}-f(p_{x})$ and $z_{4}(p_{x})=f^{'}(p_{x})\mathrm{P_{c}}$.

Denote $\widetilde{p}_{w=1}^{[2]}$ such that $L_{[2]}^{'}(\widetilde{p}_{w=1}^{[2]})=0$
and $z_{3}(\widetilde{p}_{w=1}^{[2]})+z_{4}(\widetilde{p}_{w=1}^{[2]})=0$,
when $w=1$. Following the same analysis of obtaining (\ref{eq:opt-power}),
the optimal solution of Problem (\ref{eq:final-p}) is given by 
\begin{equation}
\mathrm{P_{\mathrm{opt}}^{[2]}}=\begin{cases}
\mathrm{P_{\mathrm{max}}}, & \textrm{if }\mathrm{P_{\mathrm{max}}}\leq\widetilde{p}_{w=1}^{[2]}\\
\mathrm{P_{\mathrm{max}}}, & \textrm{if }\mathrm{P_{\mathrm{max}}}>\widetilde{p}_{w=1}^{[2]}\textrm{ and }0\leq w<\frac{-z_{4}(\mathrm{P_{\mathrm{max}}})}{z_{3}(\mathrm{P_{\mathrm{max}}})}\\
\widetilde{p}^{[2]}, & \textrm{if }\mathrm{P_{\mathrm{max}}}>\widetilde{p}_{w=1}^{[2]}\textrm{ and }\frac{-z_{4}(\mathrm{P_{\mathrm{max}}})}{z_{3}(\mathrm{P_{\mathrm{max}}})}\leq w\leq1,
\end{cases}\label{eq:opt-power-1}
\end{equation}
where $\widetilde{p}^{[2]}$ is the solution of $L_{[2]}^{'}(\widetilde{p}^{[2]})=0$
when $\textrm{if }\mathrm{P_{\mathrm{max}}}>\widetilde{p}_{w=1}^{[2]}\textrm{ and }\frac{-z_{4}(\mathrm{P_{\mathrm{max}}})}{z_{3}(\mathrm{P_{\mathrm{max}}})}\leq w\leq1$.
$\widetilde{p}_{w=1}^{[2]}$ and $\widetilde{p}^{[2]}$ can be obtained
by ``solve'' function in MATLAB.

For $K$-user system, following the same method of
obtaining Problem (\ref{RS-tradeoff-SOO-problem-2}) and (\ref{RS-tradeoff-SOO-problem-3}),
Problem (\ref{RS-tradeoff-weighted-power-problem}) can also be equivalently
approximated by a SCA problem. Specifically, let a new variable $\hat{\eta}$
represent the weighted-power EE objective value in Problem (\ref{RS-tradeoff-weighted-power-problem})
and take the place of $\eta$ in the constraint (\ref{eq:const-2-1}),
i.e., 
\begin{equation}
\hat{\eta}\leq\phi^{(n)}(x,y).\label{eq:weight-2}
\end{equation}

Then constraint (\ref{eq:tradeoff-1-c-2}) is also replaced by 
\begin{equation}
w(||\mathbf{F}||_{F}^{2}+\chi\sum_{i\in\mathcal{K}_{\mathrm{c}}}r_{i})+\mathrm{P_{c}}\leq y.\label{eq:weight-1}
\end{equation}

In summary, Problem (\ref{RS-tradeoff-weighted-power-problem}) can
be approximated by LB I as an SOCP given by \begin{subequations}\label{RS-tradeoff-MOO-problem-1}
\begin{align}
\mathop{\max}\limits _{\mathbf{f}\in\mathcal{S},\eta,\mathbf{r},x,y} & \;\;\hat{\eta}\label{RS-tradeoff-SOO-1-2-1}\\
\textrm{s.t.}\ \ \  & (\ref{eq:tradeoff-1-c-1}),(\ref{eq:const-2-2}),(\ref{eq:const-2-3})\text{,(\ref{eq:weight-2})},(\ref{eq:weight-1}),\nonumber 
\end{align}
\end{subequations}or by using LB II as a GCP given by \begin{subequations}\label{RS-tradeoff-MOO-problem-1-1}
\begin{align}
\mathop{\max}\limits _{\mathbf{f}\in\mathcal{S},\eta,\mathbf{r},x,y,\boldsymbol{\gamma},\boldsymbol{\gamma}_{\mathrm{c}}} & \;\;\hat{\eta}\label{RS-tradeoff-SOO-1-2-1-1}\\
\textrm{s.t.}\ \ \  & (\ref{eq:tradeoff-1-c-1}),(\ref{eq:low-1}),(\ref{eq:low-2}),(\ref{eq:low-5}),(\ref{eq:low-6})\text{,(\ref{eq:weight-2})},(\ref{eq:weight-1}).\nonumber 
\end{align}
\end{subequations}

\textbf{Remark 3}: Tow different metrics are proposed to investigate
the SE-EE tradeoff. The weighed-sum approach is very intuitive, because
it directly studies the MOO problem (\ref{RS-tradeoff-MOO}) with
EE metric and SE metric, and then solves the MOO problem through its
corresponding SOO problem. The weighted-power approach is an indirect
way. By controlling the proportion of power consumption in the denominator
of the objective function in (\ref{RS-tradeoff-weighted-power-problem}),
we indirectly control the proportions of EE metric and SE metric in
the original MOO problem (\ref{RS-tradeoff-MOO}).

\textbf{Remark 4}: SE Problem (\ref{RS-SE-problem}) and EE Problem
(\ref{RS-EE-problem}) are special cases of SE-EE tradeoff Problem
(\ref{RS-tradeoff-SOO}) and (\ref{RS-tradeoff-weighted-power-problem}).
In particular, when $w=0$, Problem (\ref{RS-tradeoff-SOO}) and (\ref{RS-tradeoff-weighted-power-problem})
reduce to Problem (\ref{RS-SE-problem}). When $w=1$, Problem (\ref{RS-tradeoff-SOO})
and (\ref{RS-tradeoff-weighted-power-problem}) boil down to Problem
(\ref{RS-EE-problem}). Therefore, the proposed Algorithm \ref{algorithm-SCA}
can be leveraged to solve individual SE and EE problems.

\section{Numerical Results and Discussions}

In this section, extensive numerical results are provided to evaluate
the effectiveness of our proposed algorithm and the SE-EE tradeoff
performance of RSMA. Without loss of generality, the BS is equipped
with $N_{t}=4$ transmit antennas, the noise power is $\sigma_{k}^{2}=-20$
dBm, the static circuit power consumption is $\mathrm{P_{\mathrm{c}}}=$5
dBW, $\chi=0.1$ W/(bit/s/Hz), and the iterative procedure of all
the algorithms considered in this section is terminated when the objective
values between two subsequent iterations is less than $10^{-6}$.
The signal-to-noise ratio (SNR) in the figures is defined as $10\log_{10}(\mathrm{P_{\mathrm{max}}}/\sigma_{k}^{2})$.
Some specific simulation parameters are given according to different
figures. According to \cite{Dai2017hybrid}, we consider the simplified
geometric channel model as 
\begin{equation}
\mathbf{h}_{k}=\nu_{k}[1,e^{j\frac{{2\uppi}}{\lambda_{\mathrm{c}}}d\cos\theta_{k}},...,e^{j\frac{{2\uppi}}{\lambda_{\mathrm{c}}}(N_{t}-1)d\cos\theta_{k}}]^{\mathrm{T}},\label{channel-model}
\end{equation}
where $\nu_{k}$ is the gain of channel and characterizes channel
disparity parameter, $\theta_{k}$ is the angle-of-departure (AoD)
from BS to user $k$ and characterizes the correlation between channels,
and the scalar $d$ is the interval of antennas and $\lambda_{\mathrm{c}}$
is the carrier wavelength. All algorithms are performed on a PC with
a 1.99 GHz i7-8550U CPU and 16 GB RAM, and all convex problems are
solved by the advanced CVX tool, i.e., MOSEK solver \cite{Mosek2018}.

\begin{table*}
\centering \caption{Comparison algorithm summary}
\begin{tabular}{|c|c|>{\centering}p{6cm}|c|}
\hline 
 & \textbf{System model}  & \textbf{Algorithm framework used to address the fractional programming}  & \textbf{Method used in the inner iteration}\tabularnewline
\hline 
\hline 
\textbf{RSMA-LC }  & \textbf{RSMA }  & \textbf{Low-complexity }  & \textbf{n/a}\tabularnewline
\hline 
\textbf{RSMA-LB I}  & RSMA  & SCA  & SOCP in (\ref{RS-tradeoff-SOO-problem-2})\tabularnewline
\hline 
\textbf{RSMA-LB II}  & RSMA  & SCA  & GCP in (\ref{RS-tradeoff-SOO-problem-3})\tabularnewline
\hline 
\textbf{RSMA-3}  & RSMA  & SCA  & SOCP approximation \cite{Tervo2018tradeoff} of GCP in (\ref{RS-tradeoff-SOO-problem-3})\tabularnewline
\hline 
\textbf{RSMA-D-MMSE}  & RSMA  & Dinkelbach's  & WMMSE \cite{Shiwen2013letter}\tabularnewline
\hline 
\textbf{SDMA-D-bisearch}  & SDMA  & Dinkelbach's  & Semi-closed form solution with bisearch \cite{Razaviyayn2014PHD}\tabularnewline
\hline 
\textbf{SDMA-LB II}  & SDMA  & SCA  & GCP in (\ref{RS-tradeoff-SOO-problem-3}) with $||\mathbf{f}_{\mathrm{c}}||^{2}=0$\tabularnewline
\hline 
\textbf{NOMA}  & NOMA  & SCA  & GCP \cite{Mao2019EE} with $||\mathbf{f}_{\mathrm{c}}||^{2}=0$\tabularnewline
\hline 
\end{tabular}
\end{table*}

\subsection{Convergence Analysis}

We note the fact that the computational complexity of Problem (\ref{RS-tradeoff-SOO-problem-2})
and Problem (\ref{RS-tradeoff-MOO-problem-1}) are the same, while
that of Problem (\ref{RS-tradeoff-SOO-problem-3}) and Problem (\ref{RS-tradeoff-MOO-problem-1-1})
are also the same, hence this subsection only investigates the convergence
performance of Problem (\ref{RS-tradeoff-SOO-problem-2}) and Problem
(\ref{RS-tradeoff-SOO-problem-3}) proposed under weighted-sum approach.

Denote the proposed low-complexity algorithm for two-user
system as ``RSMA-LC''. Then, denote the proposed Algorithm \ref{algorithm-SCA}
running with (\ref{RS-tradeoff-SOO-problem-2}) and (\ref{RS-tradeoff-SOO-problem-3})
as ``RSMA-LB I'' and ``RSMA-LB II'', respectively. As a comparison,
an SOCP approximation of Problem (\ref{RS-tradeoff-SOO-problem-3})
has been considered by replacing the exponential cone (\ref{eq:low-1})
and (\ref{eq:low-2}) with their convex approximations (see (43) in
\cite{Tervo2018tradeoff}) and we denote this SOCP approximation method
as ``RSMA-3''. In addition, the Dinkelbach's algorithm proposed
in \cite{Shiwen2013letter} is also considered as our benchmark algorithm,
which is denoted as ``RSMA-D-MMSE''. Basically, the idea of this
benchmark algorithm is to use the WMMSE method to solve the parametric
subproblems obtained from applying the Dinkelbach's algorithm to Problem
(\ref{RS-tradeoff-SOO}). For completeness of the analysis, the SDMA
is also considered, whose design problem can be solved by Dinkelbach's
algorithm with semi-closed form solution \cite{Razaviyayn2014PHD}
in each iteration or by (\ref{RS-tradeoff-SOO-problem-3}) with minor
modifications. Those methods are represented by ``SDMA-D-bisearch''
and ``SDMA-LB II'', respectively. The algorithms are summarized
and compared in Table I.

\begin{figure}
\centering \includegraphics[width=3.5in,height=2.6in]{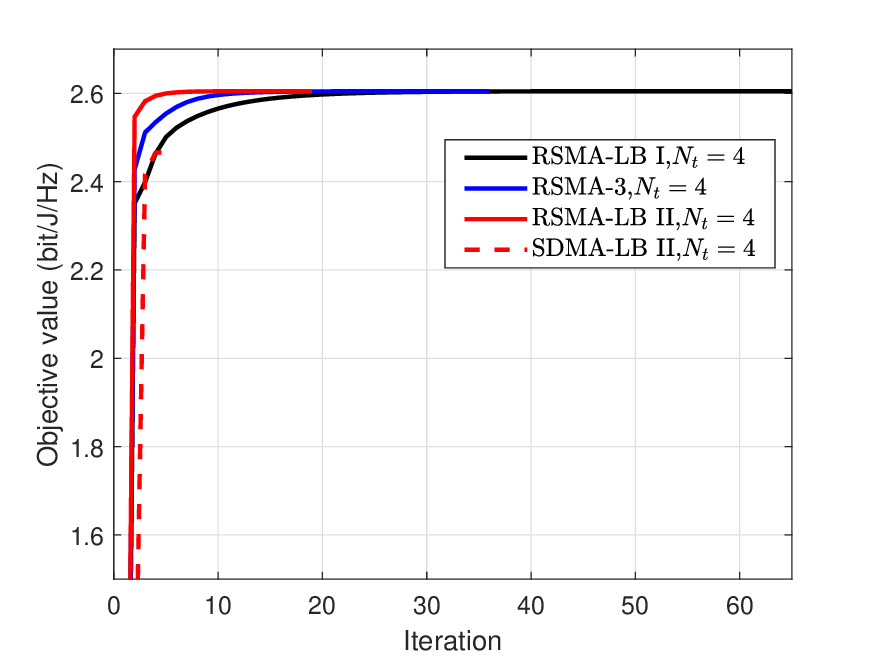}
\caption{Objective value versus number of iteration for RSMA/SDMA precoder
designed by different SCA algorithms with SNR=20 dB, $N_{t}=4$, $K=3$
and $w=0.5$.}
\label{Iteration} 
\end{figure}

\begin{figure}
\centering \includegraphics[width=3.5in,height=2.6in]{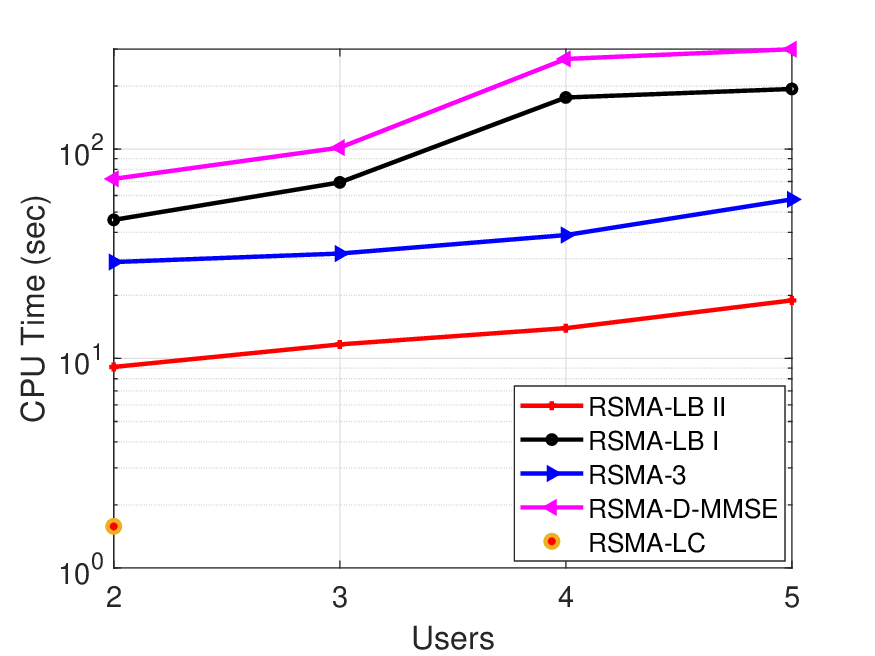}
\caption{CPU time versus $K$ by applying different algorithms
with SNR=20 dB, $N_{t}=4$ and $w=0.5$.}
\label{CPU-time} 
\end{figure}

Fig.~\ref{Iteration} compares the numbers of iterations for those
SCA algorithms with one layer iteration to converge when there are
3 users with $[\theta_{1},\theta_{2},\theta_{3}]=[0,\pi/9,2\pi/9]$
and $\nu_{1}=\nu_{2}=\nu_{3}=1$. The SNR is 20 dB and the objective
value in Y-axis is the optimal objective value of Problem (\ref{RS-tradeoff-SOO}).
It is observed that the RSMA-LB II method takes the least number of
iterations to converge since the LB II approximation is a tighter
approximation of the original Problem (\ref{RS-tradeoff-SOO}) than
LB I. But all optimization algorithms we proposed for RSMA converge
to the same boundary point, which is higher than that of the SDMA.

Fig.~\ref{CPU-time} then compares the overall CPU time of different
algorithms when there are 5 users under random channel realizations
with the entries following i.i.d. CSCG distribution. We
know that the CPU time of the proposed low-complexity algorithm is
the lowest when $K=2$ due to the closed-form solution is obtained
without iteration. When $K>2$, the proposed RSMA-LB II has the lowest
complexity among all algorithms used for RSMA optimization.

\subsection{Energy Efficiency and Spectral Efficiency Performance of the RSMA}

\begin{figure}
\centering \includegraphics[width=3.5in,height=2.6in]{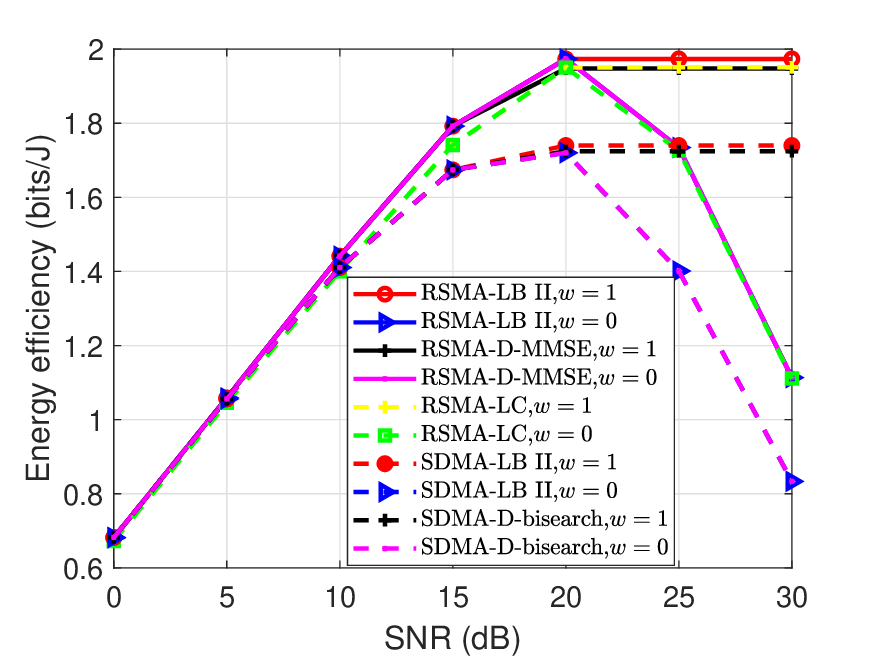} \caption{Energy efficiency versus SNR for RSMA/SDMA precoder
designed with $N_{t}=4$ and $K=2$.}
\label{EE-SNR} 
\end{figure}

\begin{figure}
\centering \subfigure[Spectral efficiency versus SNR]{ %
\begin{minipage}[t]{0.495\linewidth}%
\centering \includegraphics[width=1.8in]{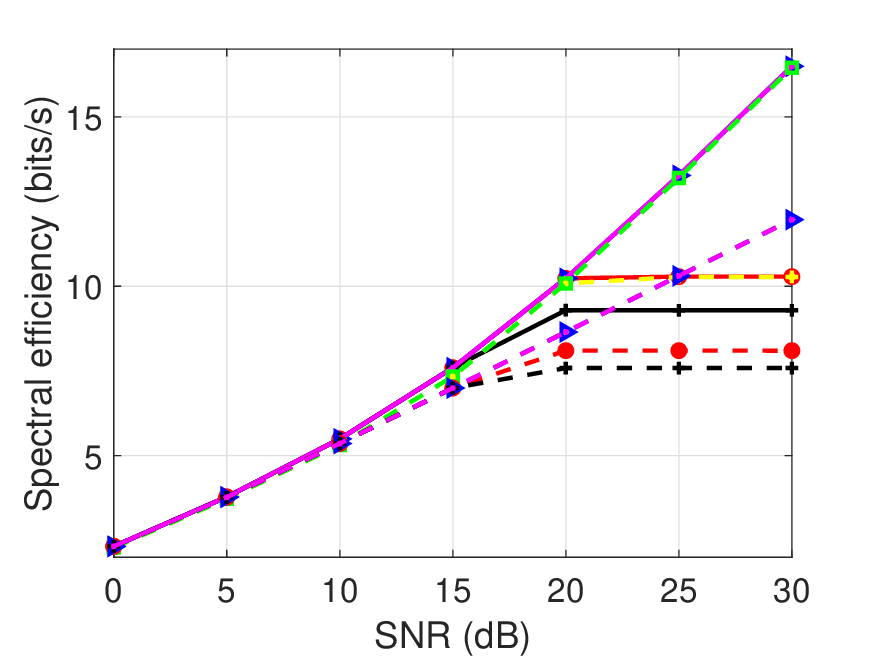} 
\end{minipage}}\subfigure[Power consumption versus SNR]{ %
\begin{minipage}[t]{0.495\linewidth}%
\centering \includegraphics[width=1.8in]{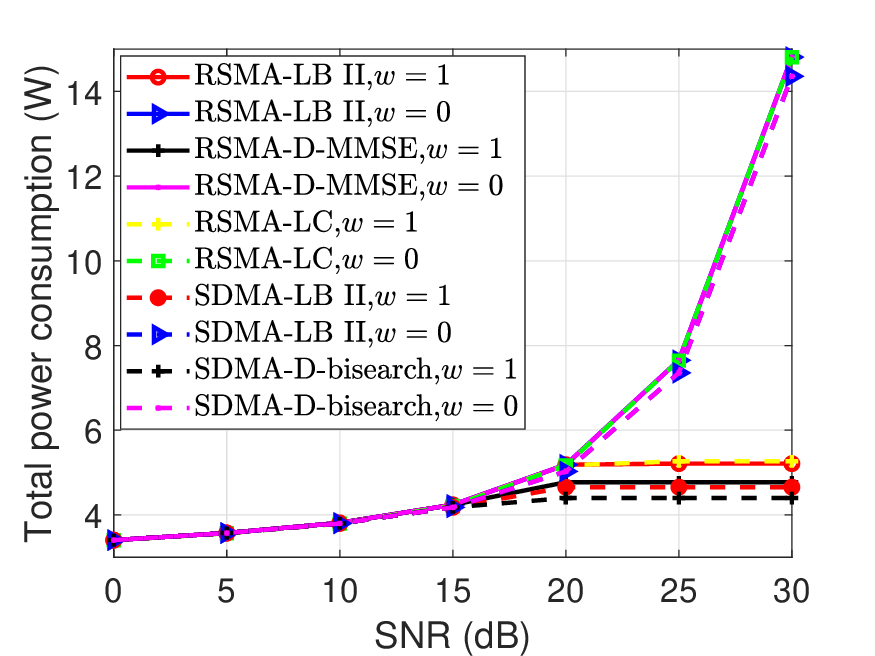} 
\end{minipage}}

\caption{Spectral efficiency and power consumption versus SNR
for RSMA/SDMA precoder designed with $N_{t}=4$ and $K=2$.}
\label{Rate-SNR} 
\end{figure}

The SE and EE benefits of RSMA have been verified in \cite{Mao2019EE}.
Specifically, \cite{Mao2019EE} separately investigates the SE maximization
problem which is solved by the WMMSE method and EE maximization problem
which is solved by the SCA method. However, how the system EE of RSMA
changes with the SNR, and how RSMA can control its SE and power consumption
to achieve high EE, have never been studied. In this subsection, we
compare the EE of RSMA with that of SDMA as SNR increases as well
as the corresponding SE and power consumption achieved with both multiple
access. As we mentioned in Remark 4, Algorithm \ref{algorithm-SCA}
can be directly adopted to solve the SE maximization Problem (\ref{RS-SE-problem})
and the EE maximization Problem (\ref{RS-EE-problem}) by setting
$w=0$ and $w=1$, respectively.

Fig.~\ref{EE-SNR} shows the EE comparison in different schemes containing
2 users with AoDs $[\theta_{1},\theta_{2}]=[0,\pi/9]$. We should
notice that when $w=0$, the value of EE is the ratio of the optimal
SE over the power consumption. For the sake of argument, the corresponding
denominators and numerators of EE, i.e., achievable SE and achievable
transmit power consumption, are also shown in Fig.~\ref{Rate-SNR}(a)
and Fig.~\ref{Rate-SNR}(b), respectively. \textit{Firstly}, the
EE performance of SCA algorithms and their corresponding Dinkelbach's
algorithms are almost the same in both RSMA and SDMA. Furthermore,
the proposed low-complexity algorithm, though with a much lower computational
complexity, achieves almost the same performance as the SCA algorithm.
\textit{Secondly}, the EE of each algorithm with $w=1$ is equal to
that with $w=0$ at low SNR (when $\mathrm{SNR}\leq\mathrm{SNR_{tradeoff}}=10\log_{10}(\widetilde{p}_{w=1}/\sigma_{k}^{2})$),
while at high SNR (when $\mathrm{SNR}>\mathrm{SNR_{tradeoff}}=10\log_{10}(\widetilde{p}_{w=1}/\sigma_{k}^{2})$)
their behaviors conflict with each other. This is because the aim
of Problem (\ref{RS-EE-problem}) is to optimize EE and keep the objective
value non-decrease, while that of Problem (\ref{RS-SE-problem}) is
to use all the available transmit power to produce maximum SE and
even sacrifice EE. \textit{Thirdly}, the EE produced by RSMA precoder
is higher than that generated by SDMA precoder. This behavior is easy
to understand in the class of '$w=0$' curves, where RSMA precoder
uses the same available power (see Fig.~\ref{Rate-SNR}(b)) to produce
higher achievable SE (see Fig.~\ref{Rate-SNR}(a)). While in the
class of '$w=1$' curves, RSMA precoder uses a little bit higher power
(see Fig.~\ref{Rate-SNR}(b)) to produce much higher SE (see Fig.~\ref{Rate-SNR}(a))
than SDMA precoder, the resulting EE value of RSMA precoder naturally
much higher than that of SDMA precoder. That is to say RSMA shows
its crucial benefits in EE communication system. Finally, the behavior
of these curves with $w=1$ in Fig.~\ref{Rate-SNR}(a) and Fig.~\ref{Rate-SNR}(b)
reveals the performance superiority of SCA algorithm over Dinkelbach's
algorithm. Basically, while achieving the same high EE, the SE resulted
by SCA algorithm is higher than that generated by Dinkelbach's algorithm,
which guarantees the QoS of the communication system.s

\subsection{The SE-EE Tradeoff of the RSMA}

This subsection investigates the SE-EE tradeoff performance of the
RSMA and SDMA where the RSMA/SDMA precoders are designed by the LB
II since it is more general for $K$-user systems and generates better
objective values than the low-complexity algorithm for the two-user
system.

\begin{figure}
\centering \includegraphics[width=3.5in,height=2.6in]{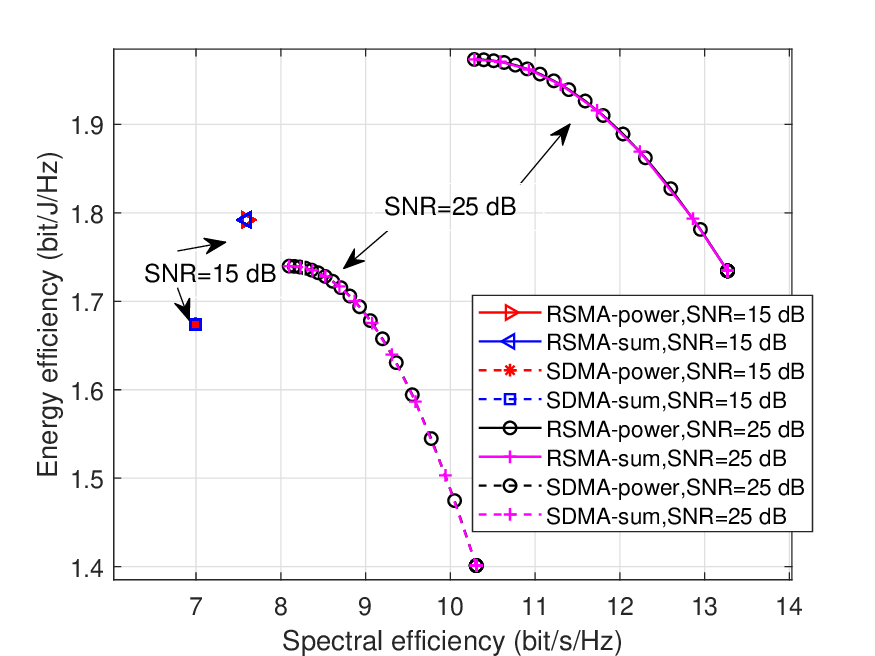}
\caption{The SE-EE tradeoff for RSMA/SDMA precoder designed by different approaches
with $N_{t}=4$, $K=2$ and $\chi=0.1$ W/(bit/s/Hz).}
\label{tradeoff-K2} 
\end{figure}

\begin{figure}
\centering \includegraphics[width=3.5in,height=2.6in]{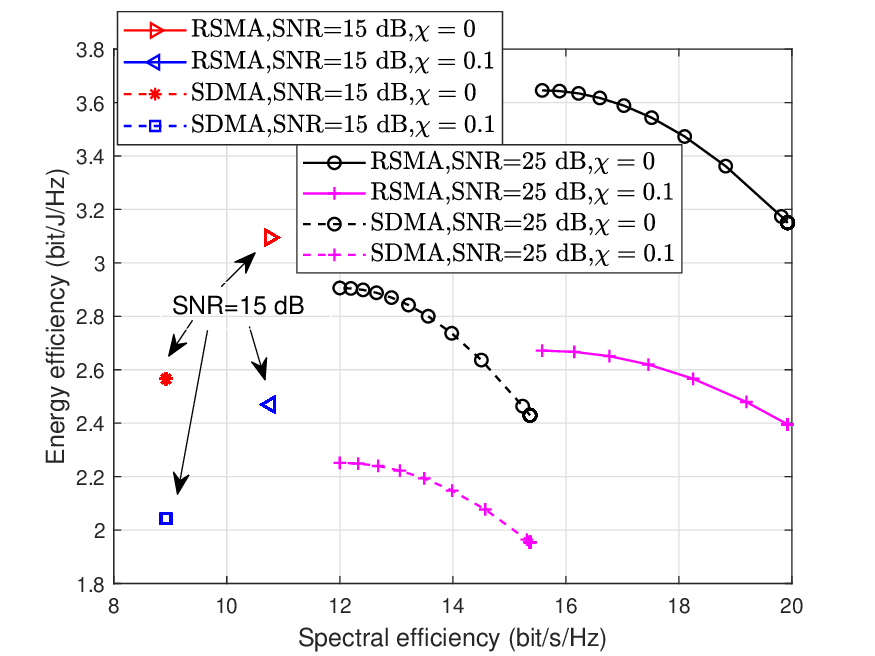}
\caption{The SE-EE tradeoff for RSMA/SDMA precoder with $N_{t}=4$ and $K=4$.}
\label{tradeoff-K4} 
\end{figure}

Fig.~\ref{tradeoff-K2} shows the SE-EE tradeoff generated by our
proposed weighted-sum approach (\ref{RS-tradeoff-SOO}) and weighted-power
approach (\ref{RS-tradeoff-weighted-power-problem}) when $K=2$ and
$[\theta_{1},\theta_{2}]=[0,\pi/9]$. \textit{Firstly}, it is obvious
that the performances of the weighted-sum and weighted-power approaches
are the same. While the mathematical model of the weighted-power metric
is more concise than the weighted-sum metric. \textit{Secondly}, when
SNR=15 dB, the EE and SE remain unchanged in the range of $w$ from
0 to 1, which indicates that the interests between EE and SE do not
conflict with each other at low SNR. \textit{Finally}, when SNR=25
dB, we noticed a tradeoff between EE and SE. Since the performance
of the weighted-sum approach and the weighted-power approach is the
same, the following simulations only adopt the weighted-sum approach
to investigate the SE-EE tradeoff performance.

Fig.~\ref{tradeoff-K4} considers the 4-user case with $[\theta_{1},\theta_{2},\theta_{3},\theta_{4}]=[0,\pi/9,2\pi/9,3\pi/9]$
at different SNR and $\chi$. It is observed that the EE decreases
with the increase of $\chi$, but the change of $\chi$ does not affect
the SE.

\begin{figure}
\centering \includegraphics[width=3.5in,height=2.6in]{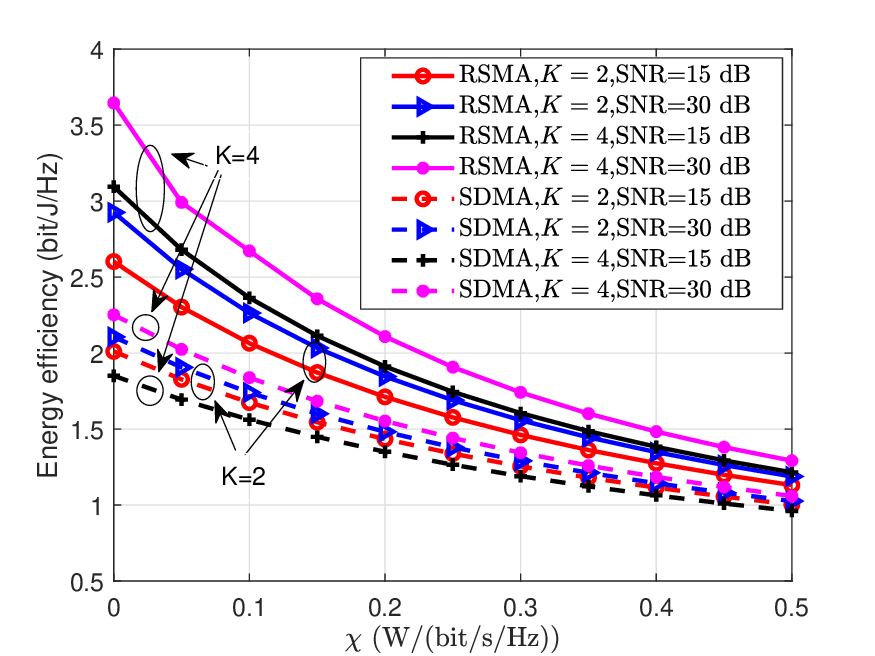}
\caption{Energy efficiency versus $\chi$ with $N_{t}=4$, $K=2$ and $w=1$.}
\label{dynamic-power} 
\end{figure}

Following the conclusion obtained from Fig.~\ref{tradeoff-K2} that
$\chi$ does not affect the SE, Fig.~\ref{dynamic-power} only depicts
the effect of rate-dependent dynamic circuit power consumption on
EE performance when $w=1$. It shows that the increment of $\chi$
leads to the decrease of the EE.

\begin{figure}
\centering \includegraphics[width=3.5in,height=2.6in]{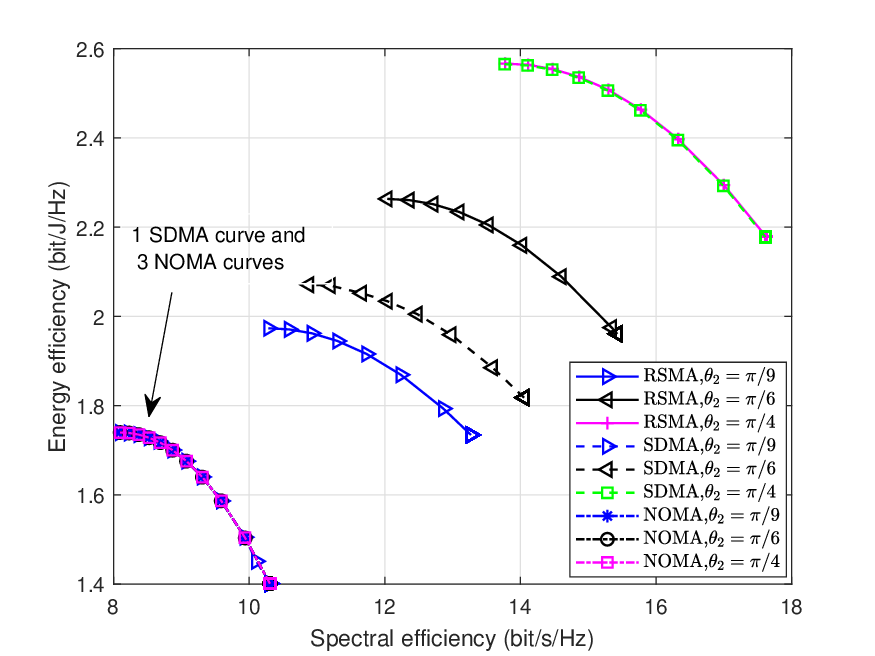}
\caption{The SE-EE tradeoff of different AoDs when SNR=25 dB, $N_{t}=4$, $K=2$
and $\chi=0.1$W/(bit/s/Hz).}
\label{tradeoff-angle} 
\end{figure}

In the following, we introduce another multiple access, namely NOMA
\cite{Mao2019EE}, as a benchmark. In the considered NOMA scenario
of $K$ users, the user with the strongest channel gain needs to decode
using SIC the messages of the remaining users before accessing its
intended stream. Fig.~\ref{tradeoff-angle} shows the SE-EE tradeoff
performance with channel gains $\{\nu_{1}=1,\nu_{2}=0.3\}$ and different
$\theta_{2}=\{\pi/9,\pi/6,\pi/4\}$. It is observed that the AoDs
has a significant effect on the performance of the RSMA, but has no
effect on that of NOMA. While as for SDMA, when users' channels are
aligned (e.g., $\theta_{2}=\{\pi/9,\pi/6\}$) and the inter-channel
interference is large, RSMA can perform more effective interference
management than SDMA. When users' channels are nearly orthogonal (e.g.,
$\theta_{2}=\pi/4$), the interference suppression capability of SDMA
is the same as that of RSMA. Generally speaking, the performance of
RSMA is better than other multiple access techniques.

\begin{figure}
\centering \includegraphics[width=3.5in,height=2.6in]{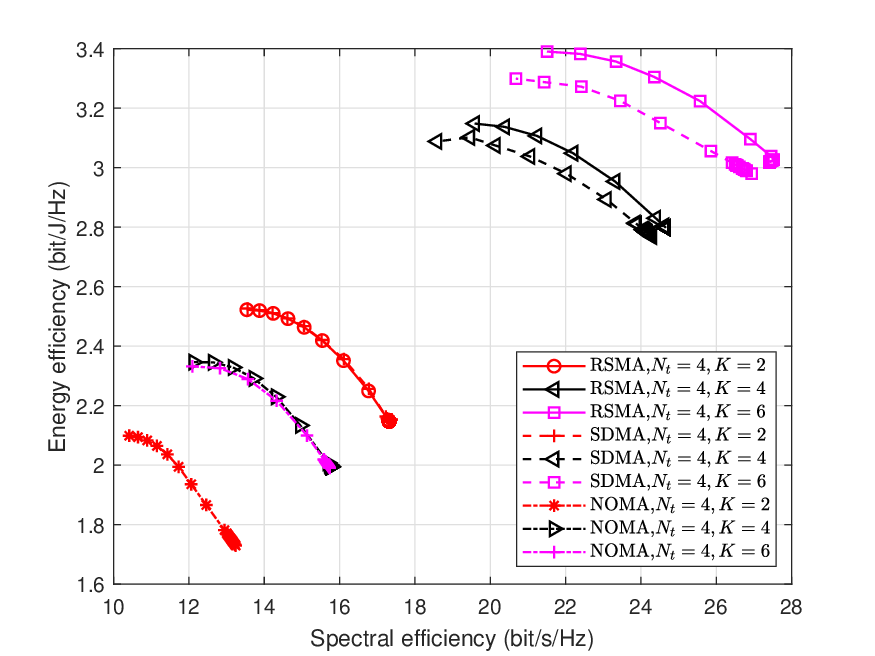}
\caption{The SE-EE tradeoff of different users when SNR=25 dB and $\chi=0.1$W/(bit/s/Hz).}
\label{tradeoff-user} 
\end{figure}

Fig.~\ref{tradeoff-user} illustrates the effect of $K$ on the SE-EE
tradeoff with RSMA, SDMA and NOMA under CSCG random channels. The
average trade-off regimes of different multiple access are generated
over 200 random channel realizations when SNR=25 dB and $\chi=0.1$W/(bit/s/Hz).
It is observed that the SE-EE trade-off regime gap between RSMA and
SDMA increases as the number of user increases. This shows that the
interference management advantage of RSMA over SDMA is more obvious
in the overloaded scenario. The performance of NOMA is the worst in
this random channel scenario.

\section{Conclusion}

In this paper, we have addressed the SE-EE tradeoff of RSMA in a multi-antenna
Broadcast Channel and have shown the potential of RSMA to boost the
SE-EE tradeoff. The tradeoff problem is a multiple-objective optimization
problem and each objective function is non-convex due to the complex
sum rate expressions and power consumption expression. In order to
overcome those challenges, we firstly proposed two approaches, namely
weighted-sum approach and weighted-power approach, to obtain the equivalent
single objective optimization problems. Then the closed-form
solution of each single objective problem is derived for the two-user
system, and a SCA algorithm is used to get the optimal precoder of
each single objective problem for the $K$-user system. Numerical
results demonstrate that the proposed low-complexity
precoding algorithm, though with a much lower computational complexity,
achieves almost the same trade-off performance as the proposed SCA
sub-optimal precoding algorithm. More importantly, compared to the
conventional SDMA and NOMA, RSMA has significant performance gains
in terms of EE and SE.

\appendices{}

\section{Proof of Lemma \ref{First-surroagte-function-lemma}}

\label{Proof-lemma-2} \textit{Proof: }We first derive the concave
lower bound function (\ref{eq:approx-2}) of common rate $R_{k}$,
and then (\ref{eq:approx-3}) could be derived following the same
procedure directly. For the convenience of deduction, (\ref{private-rate-original})
needs to be reformulated in a more readable equivalent, as 
\begin{align}
R_{k} & =\log_{2}\left(1+\frac{\mathbf{h}_{k}^{\mathrm{H}}\mathbf{f}_{k}\mathbf{f}_{k}^{\mathrm{H}}\mathbf{h}_{k}}{\sigma_{k}^{2}+\sum_{i\in\mathcal{K}\backslash\{k\}}\mathbf{h}_{k}^{\mathrm{H}}\mathbf{f}_{i}\mathbf{f}_{i}^{\mathrm{H}}\mathbf{h}_{k}}\right)\nonumber \\
 & =\log_{2}\left(1+I_{-k}^{-1}\mathbf{f}_{k}^{\mathrm{H}}\mathbf{h}_{k}\mathbf{h}_{k}^{\mathrm{H}}\mathbf{f}_{k}\right)\nonumber \\
 & =-\log_{2}\left(1-I_{k}^{-1}\mathbf{f}_{k}^{\mathrm{H}}\mathbf{h}_{k}\mathbf{h}_{k}^{\mathrm{H}}\mathbf{f}_{k}\right)\label{eq:commom-rate-1}
\end{align}
where $I_{-k}=\sigma_{k}^{2}+\sum_{i\in\mathcal{K}\backslash\{k\}}\mathbf{h}_{k}^{\mathrm{H}}\mathbf{f}_{i}\mathbf{f}_{i}^{\mathrm{H}}\mathbf{h}_{k}$
and $I_{k}=I_{-k}+\mathbf{h}_{k}^{\mathrm{H}}\mathbf{f}_{k}\mathbf{f}_{k}^{\mathrm{H}}\mathbf{h}_{k}$.
The third equation of (\ref{eq:commom-rate-1}) follows from applying
the Woodbury matrix identity \cite{book-MIL}. Let $R_{k}(\mathbf{f}_{k},I_{k})$
represent the third equation of (\ref{eq:commom-rate-1}) with constraint
$I_{k}=\sigma_{k}^{2}+\sum_{i\in\mathcal{K}}\mathbf{h}_{k}^{\mathrm{H}}\mathbf{f}_{i}\mathbf{f}_{i}^{\mathrm{H}}\mathbf{h}_{k}$,
$-\log_{2}\left(\cdot\right)$ is convex and $1-I_{k}^{-1}\mathbf{f}_{k}^{\mathrm{H}}\mathbf{h}_{k}\mathbf{h}_{k}^{\mathrm{H}}\mathbf{f}_{k}$
is jointly concave in $(\mathbf{f}_{k},I_{k})$, thus $R_{k}(\mathbf{f}_{k},I_{k})$
is jointly convex in $(\mathbf{f}_{k},I_{k})$ \cite{book-convex}
and minorized by its first-order approximation at fixed point $(\mathbf{f}_{k}^{(n)},I_{k}^{(n)})$.
Specifically, 
\begin{align}
 & R_{k}\left(\mathbf{f}_{k},I_{k}\right)\nonumber \\
 & \geq R_{k}\left(\mathbf{f}_{k}^{(n)},I_{k}^{(n)}\right)+\left(\frac{\partial R_{k}}{\partial\mathbf{f}_{k}}|_{\mathbf{f}_{k}=\mathbf{f}_{k}^{(n)}}\right)^{\mathrm{T}}\left(\mathbf{f}_{k}-\mathbf{f}_{k}^{(n)}\right)\nonumber \\
 & \thinspace\thinspace\thinspace\thinspace+\left(\frac{\partial R_{k}}{\partial\mathbf{f}_{k}^{*}}|_{\mathbf{f}_{k}^{*}=\mathbf{f}_{k}^{(n),*}}\right)^{\mathrm{T}}\left(\mathbf{f}_{k}^{*}-\mathbf{f}_{k}^{(n),*}\right)\nonumber \\
 & \thinspace\thinspace\thinspace\thinspace+\frac{\partial R_{k}}{\partial I_{k}}|_{I_{k}=I_{k}^{(n)}}\left(I_{k}-I_{k}^{(n)}\right)\nonumber \\
 & =R_{k}\left(\mathbf{f}_{k}^{(n)},I_{k}^{(n)}\right)+2\text{Re}\left\{ a_{k}\mathbf{b}_{k}^{\mathrm{H}}(\mathbf{f}_{k}-\mathbf{f}_{k}^{(n)})\right\} \nonumber \\
 & \ \ \ -a_{k}(I_{k}^{(n)})^{-2}\mathbf{f}_{k}^{(n),\mathrm{H}}\mathbf{h}_{k}\mathbf{h}_{k}^{\mathrm{H}}\mathbf{f}_{k}^{(n)}(I_{k}-I_{k}^{(n)}),\label{eq:com-approx-1}
\end{align}
where $a_{k}$ and $\mathbf{b}_{k}$ are defined in (\ref{eq:ak})
and (\ref{eq:bk}), respectively. Undo $I_{k}=\sigma_{k}^{2}+\sum_{i\in\mathcal{K}}\mathbf{h}_{k}^{\mathrm{H}}\mathbf{f}_{i}\mathbf{f}_{i}^{\mathrm{H}}\mathbf{h}_{k}$
and $I_{k}^{n}=\sigma_{k}^{2}+\sum_{i\in\mathcal{K}}\mathbf{h}_{k}^{\mathrm{H}}\mathbf{f}_{i}^{(n)}\mathbf{f}_{i}^{(n),\mathrm{H}}\mathbf{h}_{k}$,
the last equation of (\ref{eq:com-approx-1}) equals (\ref{eq:approx-2}).

Hence, the proof is completed.

\section{Proof of Proposition \ref{Theorem-socp-kkt}\label{subsec:The-proof-of-5}}

In the following, we prove that $\mathbf{F}^{o}$ is the KKT point
based on the fact that all the globally optimal solutions of a convex
optimization problem should satisfy the KKT optimality conditions
\cite{book-convex}.

Since (\ref{RS-tradeoff-SOO-problem-2}) and (\ref{RS-tradeoff-SOO-problem-3})
are both convex problems derived from SCA method, the optimal solutions
of these two problems have the same property. Here, we only prove
in detail that the optimal solution of Problem (\ref{RS-tradeoff-SOO-problem-2})
is the KKT optimality for Problem (\ref{RS-tradeoff-SOO}).

Firstly, denote by $\ensuremath{\{\mathbf{F}}^{o},\eta^{o},\mathbf{r}^{o},x^{o},y^{o}\}$
the converged solutions of Problem (\ref{RS-tradeoff-SOO-problem-2})
whose Lagrangian is then given by 
\begin{align*}
 & \mathcal{L}(\ensuremath{\mathbf{F}},\eta,\mathbf{r},x,y,\ensuremath{\boldsymbol{\lambda}})\\
= & w\eta+\frac{(1-w)}{\mathrm{P_{c}}}\sum_{i\in\mathcal{K}_{\mathrm{c}}}r_{i}-\sum_{k\in\mathcal{K}}\lambda_{k}^{(1)}(r_{k}-R_{k}^{(n)}\left(\mathbf{F}|\mathbf{F}^{o}\right))\\
 & -\sum_{k\in\mathcal{K}}\lambda_{k}^{(2)}(r_{\mathrm{c}}-R_{\mathrm{c},k}^{(n)}\left(\mathbf{F}|\mathbf{F}^{o}\right))-\lambda^{(3)}(\eta-\phi^{(n)}(x,y|x^{o},y^{o}))\\
 & -\lambda^{(4)}(x^{2}-\sum_{i\in\mathcal{K}_{\mathrm{c}}}r_{i})-\lambda^{(5)}(||\mathbf{F}||_{F}^{2}+\mathrm{P_{c}}+\chi\sum_{i\in\mathcal{K}_{\mathrm{c}}}r_{i}-y),
\end{align*}
where $\boldsymbol{\lambda}=[\lambda_{1}^{(1)},...,\lambda_{K}^{(1)},\lambda_{1}^{(2)},...,\lambda_{K}^{(2)},\lambda^{(3)},\lambda^{(4)},\lambda^{(5)}]^{\mathrm{T}}$
are the dual variables. There must exist a $\boldsymbol{\lambda}^{o}$
satisfying the following partial KKT conditions: \begin{subequations}\label{derivatives-f-1}
\begin{align}
 & \nabla_{\mathbf{F}^{*}}\mathcal{L}=\sum_{k\in\mathcal{K}}\lambda_{k}^{(1),o}\nabla_{\mathbf{F}^{*}}R_{k}^{(n)}\left(\mathbf{F}|\mathbf{F}^{o}\right)|_{\mathbf{F}=\mathbf{F}^{o}}\nonumber \\
 & +\sum_{k\in\mathcal{K}}\lambda_{k}^{(2),o}\nabla_{\mathbf{F}^{*}}R_{\mathrm{c},k}^{(n)}\left(\mathbf{F}|\mathbf{F}^{o}\right)|_{\mathbf{F}=\mathbf{F}^{o}}-\lambda^{(5),o}\mathbf{F}^{o}=\mathbf{0},\label{eq:KKT-f-1}\\
 & \nabla_{x}\mathcal{L}=\lambda^{(3),o}\frac{2x^{o}}{y^{o}}-2\lambda^{(4),o}x^{o}=0,\label{eq:f}\\
 & \nabla_{y}\mathcal{L}=\lambda^{(5),o}-\lambda^{(3),o}\left(\frac{x^{o}}{y^{o}}\right)^{2}=0,\label{eq:c}\\
 & \lambda_{k}^{(1),o}(r_{k}^{o}-R_{k}^{(n)}\left(\mathbf{F}^{o}|\mathbf{F}^{o}\right))=0,\forall k\in\mathcal{K},\label{eq:KKT-f-2}\\
 & \lambda_{k}^{(2),o}(r_{\mathrm{c}}^{o}-R_{\mathrm{c},k}^{(n)}\left(\mathbf{F}^{o}|\mathbf{F}^{o}\right))=0,\forall k\in\mathcal{K},\label{eq:KKT-3}\\
 & \lambda^{(3),o}(\eta-\phi^{(n)}(x^{o},y^{o}))=0,\label{eq:v}\\
 & \lambda^{(5),o}(||\mathbf{F}^{o}||_{F}^{2}+\mathrm{P_{c}}+\chi\sum_{i\in\mathcal{K}_{\mathrm{c}}}r_{i}^{o}-y^{o})=0.\label{eq:KKT-f-3}
\end{align}
\end{subequations}

It is straightforward that\begin{subequations}\label{derivatives-f-2}
\begin{align}
R_{k}^{(n)}\left(\mathbf{F}^{o}|\mathbf{F}^{o}\right) & =R_{k}\left(\mathbf{F}^{o}\right),\label{eq:A1-f-1}\\
R_{\mathrm{c},k}^{(n)}\left(\mathbf{F}^{o}|\mathbf{F}^{o}\right) & =R_{\mathrm{c},k}\left(\mathbf{F}^{o}\right).\label{eq:fff}
\end{align}
\end{subequations}

Furthermore, the first-order partial derivatives with respect to $\mathbf{F}$
for $R_{k}\left(\mathbf{F}\right)$ and $R_{k}^{(n)}\left(\mathbf{F}|\mathbf{F}^{o}\right)$
are given by\begin{subequations}\label{derivatives-f} 
\begin{align}
\nabla_{\mathbf{f}_{k}^{*}}R_{k}\left(\mathbf{F}\right)|_{\mathbf{f}_{k}^{*}=\mathbf{f}_{k}^{*,o}} & =\frac{\mathbf{h}_{k}^{\mathrm{H}}\mathbf{f}_{k}^{o}\mathbf{h}_{k}}{I_{k}^{o}}=\nabla_{\mathbf{f}_{k}^{*}}R_{k}^{(n)}\left(\mathbf{F}|\mathbf{F}^{o}\right)|_{\mathbf{f}_{k}^{*}=\mathbf{f}_{k}^{*,o}},\\
\nabla_{\mathbf{f}_{i}^{*}}R_{k}\left(\mathbf{F}\right)|_{\mathbf{f}_{i}^{*}=\mathbf{f}_{i}^{*,o}} & =-\frac{\mathbf{h}_{k}^{\mathrm{H}}\mathbf{f}_{k}^{o}\mathbf{f}_{k}^{\mathrm{H},o}\mathbf{h}_{k}\mathbf{h}_{k}^{\mathrm{H}}\mathbf{f}_{i}^{o}\mathbf{h}_{k}}{I_{k}^{o}I_{-k}^{o}}\nonumber \\
 & =\nabla_{\mathbf{f}_{i}^{*}}R_{k}^{(n)}\left(\mathbf{F}|\mathbf{F}^{o}\right)|_{\mathbf{f}_{i}^{*}=\mathbf{f}_{i}^{*,o}},
\end{align}
\end{subequations}where $I_{-k}^{o}=\sigma_{k}^{2}+\sum_{i\in\mathcal{K}\backslash\{k\}}\mathbf{h}_{k}^{\mathrm{H}}\mathbf{f}_{i}^{o}\mathbf{f}_{i}^{\mathrm{H},o}\mathbf{h}_{k}$
and $I_{k}^{o}=I_{-k}^{o}+\mathbf{h}_{k}^{\mathrm{H}}\mathbf{f}_{k}^{o}\mathbf{f}_{k}^{\mathrm{H},o}\mathbf{h}_{k}$.
With (\ref{derivatives-f}), we arrive at 
\begin{align}
\nabla_{\mathbf{F}}R_{k}\left(\mathbf{F}\right)|_{\mathbf{F}=\mathbf{F}^{o}} & =\nabla_{\mathbf{F}}R_{k}^{(n)}\left(\mathbf{F}|\mathbf{F}^{o}\right)|_{\mathbf{F}=\mathbf{F}^{o}}.\label{eq:A3-f}
\end{align}
Then, the equivalence relation between two first-order partial derivatives
of $R_{\mathrm{c},k}\left(\mathbf{F}\right)$ and $R_{\mathrm{c},k}^{(n)}\left(\mathbf{F}|\mathbf{F}^{o}\right)$
is directly given by 
\begin{align}
\nabla_{\mathbf{F}}R_{\mathrm{c},k}\left(\mathbf{F}\right)|_{\mathbf{F}=\mathbf{F}^{o}} & =\nabla_{\mathbf{F}}R_{\mathrm{c},k}^{(n)}\left(\mathbf{F}|\mathbf{F}^{o}\right)|_{\mathbf{F}=\mathbf{F}^{o}}.\label{eq:A3-f-1}
\end{align}

By substituting (\ref{eq:A3-f}) and (\ref{eq:A3-f-1}) into (\ref{eq:KKT-f-1}),
and substituting (\ref{eq:A1-f-1}) and (\ref{eq:fff}) into (\ref{eq:KKT-f-2})
and (\ref{eq:KKT-3}) respectively, we obtain \begin{subequations}\label{derivatives-f-3}
\begin{align}
 & \sum_{k\in\mathcal{K}}\lambda_{k}^{(1),o}\nabla_{\mathbf{F}^{*}}R_{k}\left(\mathbf{F}\right)|_{\mathbf{F}=\mathbf{F}^{o}}\nonumber \\
 & +\sum_{k\in\mathcal{K}}\lambda_{k}^{(2),o}\nabla_{\mathbf{F}^{*}}R_{\mathrm{c},k}\left(\mathbf{F}\right)|_{\mathbf{F}=\mathbf{F}^{o}}-\lambda^{(5),o}\mathbf{F}^{o}=\mathbf{0},\label{eq:KKT-f-1-1}\\
 & \lambda_{k}^{(1),o}(r_{k}^{o}-R_{k}\left(\mathbf{F}^{o}\right))=0,\forall k\in\mathcal{K},\label{eq:KKT-f-2-1}\\
 & \lambda_{k}^{(2),o}(r_{\mathrm{c}}^{o}-R_{\mathrm{c},k}\left(\mathbf{F}^{o}\right))=0,\forall k\in\mathcal{K}
\end{align}
\end{subequations}

Now, it can be readily verified that the set of equations (\ref{eq:f}),
(\ref{eq:c}), (\ref{eq:v}), (\ref{eq:KKT-f-3}), and (\ref{derivatives-f-3})
constitute exactly the KKT conditions of Problem (\ref{RS-tradeoff-SOO-problem-1}).
Since Problem (\ref{RS-tradeoff-SOO-problem-1}) is equivalent to
Problem (\ref{RS-tradeoff-SOO}), $\mathbf{F}^{o}$ is the KKT optimality
for Problem (\ref{RS-tradeoff-SOO}).

Hence, the proof is completed.


 \bibliographystyle{IEEEtran}
\bibliography{bibfile}

\end{document}